\documentclass[preprint,11pt,sort&compress]{elsarticle}
\usepackage[T1]{fontenc}
\usepackage{lmodern}
\usepackage{amsmath,amssymb,amsthm,mathtools,booktabs,array,graphicx,enumitem}
\usepackage{geometry}
\usepackage[hidelinks]{hyperref}
\newtheorem{theorem}{Theorem}

\newtheorem{corollary}[theorem]{Corollary}
\theoremstyle{remark}

\newcommand{\E}{\mathbb E}
\newcommand{\Prob}{\mathbb P}

\newcommand{\kl}{\operatorname{kl}}

\journal{Journal of Complexity}
\hypersetup{pdfauthor={Yujiao Peng and Qifan Yang}}
\hypersetup{pdftitle={Feedback Rounds and Query Costs in Sparse Product Pure Exploration}}
\begin{document}
\raggedbottom
\begin{frontmatter}
\title{Feedback Rounds and Query Costs in Sparse Product Pure Exploration}
\author[sysu]{Yujiao Peng\fnref{equal}}
\author[ruc]{Qifan Yang\corref{cor}\fnref{equal}}
\ead{Y2771433453@ruc.edu.cn}
\address[sysu]{Sun Yat-sen University, China}
\address[ruc]{School of Business, Renmin University of China, Beijing, China}
\cortext[cor]{Corresponding author.}
\fntext[equal]{Yujiao Peng and Qifan Yang contributed equally to this work.}
\begin{abstract}
How much can one more round of feedback save when a measurement combines several settings into a single noisy reading? We study a product decision with $K$ blocks and $m$ nonbaseline labels per block. At most $s$ blocks have nonzero effects, of arbitrary sign. Each query tests labels in at most $b$ blocks. The final assignment must have total loss at most $\varepsilon$ with probability at least $1-\delta$. On the benchmark $K=s^2$, $1\le b\le s$ and $s^\nu\le m\le s^\zeta$ for fixed $0<\nu\le1\le\zeta$, we characterize two minimax costs for sufficiently large $s$. The optimal orders are $(Km/b)\chi U_B$ readings and $Km\chi U_B$ settings for $B\ge1$ batches, where $\chi=\sigma^2s^2/\varepsilon^2\ge1$ for noise proxy $\sigma^2$ and $\delta\le1/16$. The factor $U_B$ is obtained by applying $B-1$ logarithms to $\log(es)$, clipping each iterate at one and imposing a floor of $\log(1/\delta)/s$. One algorithm attains both bounds. For general dimensions, a polynomial-time algorithm attains the optimal setting cost. When $K=s^2$, $m=b=s$ and $\log(1/\delta)=s$, both fully adaptive orders are attainable in polynomial real-arithmetic time when $\chi=\Omega((s/\log(es))^{1/3})$. With identical budgets and decoder, four batches succeeded on 1,070 of 1,200 paired simulated instances and two batches on 779.
\end{abstract}
\end{frontmatter}
\newcommand{\MainOneA}{526}
\newcommand{\MainOneB}{1,029}
\newcommand{\MainOneDelta}{-24.56}
\newcommand{\MainOneLo}{-29.14}
\newcommand{\MainOneHi}{-19.85}
\newcommand{\MainTwoA}{557}
\newcommand{\MainTwoB}{530}
\newcommand{\MainTwoDelta}{1.32}
\newcommand{\MainTwoLo}{-2.45}
\newcommand{\MainTwoHi}{5.08}
\newcommand{\MainThreeA}{445}
\newcommand{\MainThreeB}{435}
\newcommand{\MainThreeDelta}{0.49}
\newcommand{\MainThreeLo}{-3.26}
\newcommand{\MainThreeHi}{4.24}
\newcommand{\MainFourA}{26}
\newcommand{\MainFourB}{1,861}
\newcommand{\MainFourDelta}{-89.60}
\newcommand{\MainFourLo}{-91.54}
\newcommand{\MainFourHi}{-87.17}
\newcommand{\EndpointZero}{1,120}
\newcommand{\EndpointFloor}{1,114}
\newcommand{\EndpointFloorAll}{2,037}
\newcommand{\EndpointTail}{2.86}
\newcommand{\EndpointRem}{11.1}
\newcommand{\DevelopCap}{4}
\newcommand{\DevelopBlocks}{1}
\newcommand{\DevelopGamma}{0.1}
\newcommand{\CoverLowHit}{171}
\newcommand{\CoverFullHit}{0}
\newcommand{\CoverShortHit}{176}
\newcommand{\MediumAdaptiveHit}{25}
\newcommand{\MediumFixedHit}{18}
\newcommand{\MediumShortHit}{5}

\section{Introduction}
\begin{figure}[htbp]
\centering
\includegraphics[width=\textwidth]{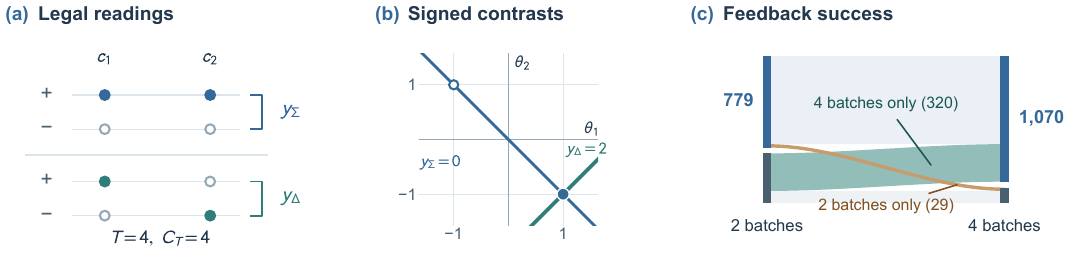}
\caption{Signed cancellation and paired feedback outcomes.
(a) Each row is a legal reading of at most two nonbaseline labels (filled circles)
in distinct blocks; subtracting the lower reading from the upper in each pair
cancels the intercept $\mu_0$.
(b) Without noise, $y_\Sigma=\theta_1+\theta_2=0$ leaves $(-1,1)$ and $(1,-1)$
indistinguishable; adding $y_\Delta=\theta_1-\theta_2=2$ identifies $(1,-1)$.
(c) Outcomes on $1{,}200$ paired signed instances ($s=m=b=64$, $\chi=4$) under
identical budgets ($T=262{,}144$, $C_T=3{,}178{,}496$) and a shared decoder.
Upper blue bars count successes ($779$ vs.\ $1{,}070$) and lower bars count
failures; ribbon widths represent pair counts, with $320$ instances succeeding
only under four batches and $29$ only under two.}
\label{fig:feedback}
\end{figure}

One noisy reading can test several adjustments at once. If two adjustments in different blocks have equal and opposite effects, their sum is zero. Keeping the positive adjustment and returning the other block to baseline still improves the final decision. A second contrast distinguishes the two assignments in Fig.~\ref{fig:feedback}a,b. More generally, pooling settings spares readings but can leave several labels consistent with the same observation. Feedback allows the next measurement to distinguish those labels.

We study this tradeoff for $K$ blocks with $m$ nonbaseline labels each. At most $s$ blocks have nonzero effects. A query changes at most $b$ blocks and returns a noisy scalar sum. The learner ultimately chooses a label for every block. We charge one reading per scalar observation and one setting for each block tested away from baseline in that observation. For example, pooling $b$ nonbaseline labels in a single measurement costs one reading and $b$ settings, even if their effects cancel.

\begin{figure}[tb]
\centering
\includegraphics[width=.67\linewidth]{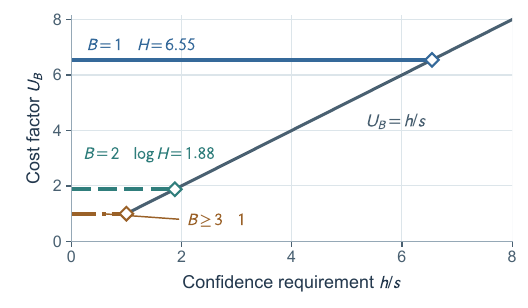}
\caption{Cost factor $U_B=\max\{L_{B-1}(H),h/s\}$ in Theorem~\ref{thm:uniform}
evaluated at $s=256$ ($H=\log(es)$). The feedback plateaus at $B=1$, $B=2$,
and $B\ge3$ have heights $H$, $\log H$, and $1$. Each open diamond marks
the point at which its plateau meets the common line $U_B=h/s$.}
\label{fig:rate}
\end{figure}

Fix the accuracy $\varepsilon$ and failure probability $\delta$. What do $B$ opportunities to plan measurements buy? Write $H=\log(es)$, $h=\log(1/\delta)$, and $\chi=\sigma^2s^2/\varepsilon^2$ for noise proxy $\sigma^2$. On the benchmark $K=s^2$, $1\le b\le s$, $s^\nu\le m\le s^\zeta$ with fixed $0<\nu\le1\le\zeta$, $\chi\ge1$ and $h\ge\log16$, Theorem~\ref{thm:uniform} gives the minimax orders
\[
  T_B^*=\Theta\!\left(\frac{Km}{b}\chi U_B\right),\qquad
  C_B^*=\Theta(Km\chi U_B),\qquad
  U_B=\max\{L_{B-1}(H),h/s\}.
\]
Here $L_j$ applies the logarithm $j$ times, clipping each iterate at one. One batch pays a factor $\max\{H,h/s\}$, and two batches pay $\max\{1,\log H,h/s\}$. Later batches remove further logarithmic layers until $\max\{1,h/s\}$ is reached (Fig.~\ref{fig:rate}). Allowing a learner to stop early along a predetermined query order still gives the one-batch cost (Corollary~\ref{thm:extremes}).

The two minimax costs are defined separately. A single construction nevertheless attains both upper bounds for every batch allowance on this benchmark. Its final step may enumerate residual supports. The computational result is more specific. At $m=b=s$ and $h=s$, a multiscale filter, followed by a certified residual decoder, attains both fully adaptive orders in polynomial real-arithmetic time and $\log^*H+O(1)$ batches whenever $\chi=\Omega((s/H)^{1/3})$ (Theorem~\ref{thm:adaptive-compute}). At $\chi=1$ the corresponding constant-factor polynomial-time guarantee for arbitrary signed effects remains open.

For arbitrary dimensions, blockwise rank elimination with signed Hadamard cycles attains the optimal setting cost in polynomial time. It also attains the matching reading order when $b=O(\chi U_B^m)$, where $U_B^m$ replaces $H$ by $\log(em)$ in the batch factor (Theorem~\ref{thm:rank-general}). At larger capacities the information carried by one scalar reading can limit the gain from pooling.

Finite budgets expose two other features of the problem. With the same budgets and decoder, four batches succeeded on $1{,}070$ of $1{,}200$ paired signed instances and two batches on $779$ (Fig.~\ref{fig:feedback}c). In a separate low-budget study, changing the decoder on identical readings raised success from $6/256$ to $214/256$. An independent study then held the decoder fixed and varied which labels the measurements visited. Section~\ref{sec:experiments} reports all three comparisons.

\subsection{Related work}
The value of adapting the information collected depends on the output being sought. For mean computation on finite-dimensional sequence spaces, Heinrich \cite{heinrich2024} determines optimal randomized errors for adaptive and nonadaptive information. Here the output is a label assignment and the accuracy is fixed in advance. We count the measurements and the individual settings needed to achieve that accuracy at a specified confidence.

Iterated logarithms and $\log^*$ round thresholds already occur in limited-round coin tossing and arm identification \cite{agarwal2017}. Tuynman and Degenne \cite{tuynman2025} give instance-dependent bounds for batched pure exploration. Du, Kuroki, and Chen \cite{du2021} study combinatorial pure exploration under full-bandit and partial linear feedback. In the full-bandit model a scalar observation is a linear sum and the objective is to identify an optimal action. Our observation may test at most $b$ blocks, while the decision assigns a label to all $K$. The two costs account for the width of each query as well as the number of observations. Maiti, Xu, and Jamieson \cite{maiti2026} compare adaptive and nonadaptive pure exploration on linear action sets, including a multitask setting. The supplement compares reading bounds when $b=K$ removes the capacity constraint. Malek, Aglietti, and Chiappa \cite{malek2023} study additive causal bandits under intervention structure and give instance-dependent elimination bounds.

Successive elimination for individually sampled arms goes back to Even-Dar,
Mannor, and Mansour \cite{evendar2006}, and Rejwan and Mansour
\cite{rejwan2020} use Hadamard designs to estimate arm means from full-bandit
sums. We pack history-dependent candidate lists into capacity-legal signed
Hadamard cycles and prove simultaneous error bounds under conditional
sub-Gaussian noise. For signed sparse recovery, our balanced layouts build on
lossless expansion and $\ell_1$ stability \cite{berinde2008,indyk2014}, while
Kamath and Price \cite{kamath2019} analyze round tradeoffs for unrestricted
linear sketches.

Constructing unbalanced expanders deterministically or with exponentially
small failure probability poses a distinct obstacle when $h=\Theta(s)$: at
$b=s$, a table of $O(sm)$ rows over $s^2m$ cells has right-to-left size ratio
$O(1/s)$. Explicit constant-degree lossless expanders handle constant
imbalance \cite{golowich2024}, and randomized constructions for polynomial
imbalance carry fixed polynomial or negligible construction-error guarantees
\cite{applebaum2023}. Our multiscale filter instead bounds collisions on the
fixed unknown instance and certifies expansion only on the small residual
allowance $2k$.

\section{Model and performance criteria}\label{sec:model}
Block $i\in[K]$ has baseline label $0$ and $m$ nonbaseline labels. An
assignment $x$ with at most $b$ nonzero coordinates is a legal query.
One physical reading returns
\begin{equation}\label{eq:observation}
 Y=\mu_0+\sum_{i=1}^K\theta_{i,x_i}+\xi,\qquad \theta_{i,0}=0 .
\end{equation}
Both $\mu_0$ and the effects $\theta_{i,a}$ are unknown. The active block set
$S=\{i:\exists a,\ \theta_{i,a}\ne0\}$ has size at most $s$, and active
entries are arbitrary finite real numbers. The learner knows
$K,m,b,s,\varepsilon,\delta$ and a valid noise proxy $\sigma^2$. Write
$\mathcal F_{t-1}$ for the history of prior physical outcomes and private
randomness drawn before query $t$, including the current batch plan; each
fresh private draw is independent of $\mathcal F_{t-1}$. Conditional noise
satisfies
\begin{equation}\label{eq:mgf}
 \E[e^{\lambda\xi_t}\mid\mathcal F_{t-1},x_t]
       \le e^{\lambda^2\sigma^2/2},\qquad \lambda\in\mathbb R.
\end{equation}
Averaging $q$ fresh query--baseline differences cancels $\mu_0$, yields noise
proxy $2\sigma^2/q$, and costs $2q$ physical readings.

The learner stops almost surely and outputs a full assignment $\widehat x$
with no capacity limit and total assignment loss
\begin{equation}\label{eq:loss}
 \mathcal R_\theta(\widehat x)=
 \sum_{i=1}^K\{\max_{0\le a\le m}\theta_{i,a}-\theta_{i,\widehat x_i}\}.
\end{equation}
An $(\varepsilon,\delta)$-PAC algorithm satisfies
$\Prob_\theta\{\mathcal R_\theta(\widehat x)\le\varepsilon\}\ge 1-\delta$
for every admissible instance and noise law. We track both the physical
reading count $T$ and the cumulative setting count
\begin{equation}\label{eq:settings}
 C_T=\sum_{t=1}^T|\{i:x_{t,i}\ne0\}|\le bT .
\end{equation}
A baseline-only measurement contributes one reading and zero settings. A block tested on two successive readings contributes a setting on each. Under each protocol, the minimax reading and setting costs are the separate
infima over $(\varepsilon,\delta)$-PAC algorithms of their worst-instance
expected values of $T$ and $C_T$. A predetermined order fixes its query
sequence from private randomness before observing data, though its stopping
time may depend on feedback. In a strict $B$-batch algorithm, all queries and
the length of each batch are fixed before its first observation; empty
batches are allowed.

Let $N=Km$ be the number of nonbaseline cells and $n_b=N/b$ the reading-count scale for one full pass. Covering every cell requires at least $\lceil n_b\rceil$ physical readings. Fix $0<\nu\le1\le\zeta$. The signed
computational benchmark is
\begin{equation}\label{eq:benchmark}
 \begin{gathered}
 K=s^2,\quad 1\le b\le s,\quad s^\nu\le m\le s^\zeta,
 \quad H=\log(es),\\
 \log16\le h=\log(1/\delta)\le s,\qquad
 \chi=\sigma^2s^2/\varepsilon^2\ge1 .
 \end{gathered}
\end{equation}
Theorem~\ref{thm:uniform} extends \eqref{eq:benchmark} to all $h\ge\log16$,
and Theorem~\ref{thm:rank-general} treats arbitrary $(K,m,b,s)$. Natural
logarithms are used unless noted otherwise. Implied constants may depend on
$\nu,\zeta$, and asymptotic bounds hold for $s\ge s_0(\nu,\zeta)$ uniformly
in the remaining parameters; below $s_0$, averaging
$\lceil16\chi\log(2N/\delta)\rceil$ single-cell pairs per cell ensures
simultaneous $\varepsilon/(2s)$ estimation error and hence regret at most
$\varepsilon$. Running-time bounds count real arithmetic operations,
comparisons, square roots (as in the cited LP solver), and index operations;
bit complexity additionally depends on input encoding lengths.

\section{Main results}\label{sec:main}
For $u\ge1$ set $L_0(u)=u$ and $L_{q+1}(u)=\max\{1,\log L_q(u)\}$.
The clipped logarithm records the point at which another round ceases
to lower the order of the query cost. We use $\log^*u$ for the number
of natural logarithms needed to reach at most one.

\begin{theorem}[Cost of confidence and feedback]\label{thm:uniform}
Retain the dimension and noise-scale conditions in \eqref{eq:benchmark}
and allow every $h\ge\log16$. Set $U_B=\max\{L_{B-1}(H),h/s\}$.
For sufficiently large $s$, the two strict $B$-batch minimax costs are
\[
  T_B^*=\Theta(n_b\chi U_B),\qquad C_B^*=\Theta(N\chi U_B).
\]
Constants and the starting scale depend only on $\nu,\zeta$.
One algorithm attains both upper bounds on every execution path.
The lower bounds hold for independent Gaussian noise with known
zero intercept.
\end{theorem}

\begin{corollary}[Endpoint protocols]\label{thm:extremes}
A predetermined order with feedback-dependent stopping pays the same
orders as one batch. Unrestricted adaptation pays the common normalized
factor $\chi R$, where $R=\max\{1,h/s\}$. For $z=\max\{2,h/s\}$ let
$j_z(H)=\min\{j\ge0:L_j(H)\le z\}$. A uniform cap of $1+j_z(H)$
batches suffices at the adaptive rate. With either adaptive cost held
within a fixed multiplier $L$, a cap of at least
$1+j_z(H)-O_{\nu,\zeta,L}(1)$ is necessary.
\end{corollary}

\begin{theorem}[Optimization at the adaptive rate]\label{thm:adaptive-compute}
Under the full benchmark \eqref{eq:benchmark}, including $h\le s$,
write $g=1+h/H$. Both optimal pathwise adaptive cost orders are
attainable in $\log^*H+O(1)$ batches in polynomial real-arithmetic
time if any of the following conditions holds:
\begin{enumerate}[label=\arabic*)]
 \item $g=O(\chi^2)$;
 \item $g=O(\chi^3)$ and $g^{2/3}b^2/N\le s^{-\gamma}$ for fixed
       $\gamma>0$;
 \item $b=O(\chi)$.
\end{enumerate}
The constants may depend on $\nu,\zeta,\gamma$ and the fixed implied
multipliers. In particular, at $m=b=s$ and $h=s$, the condition
$\chi=\Omega((s/H)^{1/3})$ suffices for unrestricted signed effects.
For every fixed $\eta>0$, the whole benchmark admits the same cost
and batch orders in
\begin{equation}\label{eq:runtime}
 \operatorname{poly}_{\nu,\zeta,\eta}(s,m,T_{\max})e^{\eta h/\chi}
\end{equation}
operations, with query constants allowed to depend on $\eta$.
Here $T_{\max}$ is the deterministic observation cap.
\end{theorem}

\begin{theorem}[Setting cost in general dimensions]\label{thm:rank-general}
For arbitrary $1\le s,b\le K$, $m\ge2$, $\chi\ge1$ and
$h\ge\log16$, put $H_m=\log(em)$ and
$U^m_B=\max\{L_{B-1}(H_m),h/s\}$. Then
\[
 C_B^*=\Theta(Km\chi U^m_B),\qquad
 T_B^*=\Omega((Km/b)\chi U^m_B).
\]
A polynomial-time algorithm with at most $B$ batches attains the
setting upper bound pathwise with $T=O((Km/b)(b+\chi U^m_B))$.
Thus it attains both optimum orders when $b=O(\chi U^m_B)$ with a
fixed multiplier. The bounds before this last condition use
absolute constants.
\end{theorem}

The two lower bounds have different sources. Under a Gaussian reference instance, a batch constraint limits how quickly the learner can concentrate settings on a hidden cell. Transferring exposure from one batch to the next yields the iterated logarithm. A separate pair of opposite-sign alternatives enforces the confidence term. Since $C_T\le bT$, both arguments give reading lower bounds as well. Section~\ref{sec:argument} describes a single measurement construction that matches both costs. Section~\ref{sec:computation} treats the cost of computing its decision.

\section{Where the logarithms enter}\label{sec:argument}
Complete proofs of the layout, conditional-noise, and failed-path bounds
appear in the supplement.

\subsection{A stopped lower bound for limited feedback}
Partition the $K=s^2$ blocks into $s$ groups of $s$. Each group is
empty with probability $1/2$, independently, and otherwise has one
uniformly chosen cell of value $8\varepsilon/s$.
On a successful path a learner misses at most $s/8$ active cells,
so a uniformly drawn group, conditional on being active, has probability
at least $5/8$ of having its true cell selected. Under the reference law $P_0$ that leaves a designated
group empty, an independently drawn uniform cell in that group is
selected with probability at most $1/m$.

Let $V=\sup_\theta\E_\theta C_T/(Km)$ under independent Gaussian
noise with $\mu_0=0$. Writing $w_j$ for the designated cell's settings
in batch $j$, summing empty-group budgets before averaging over cells
gives $P_0\{w_j\ge u\}\le2V/u$ and total reference exposure
$\E_0[\sum_j w_j]\le2V$. By the stopped Gaussian likelihood identity,
$\mathrm{KL}(P_0\Vert P_+)\le64V/\chi$, where $P_+$ activates the
designated cell; Pinsker's inequality then forces $V=\Omega(\chi)$.

Batch limits enter when transferring the planned exposure $w_{j+1}$
from $P_0$ to $P_+$. Set $W_j=\sum_{i\le j}w_i$ and let $\Lambda_j$
be the likelihood ratio through batch $j$. The exponential Gaussian
martingale gives
\[
 P_+\{W_j\le\chi t,\ \log\Lambda_j>64t+u\}\le e^{-u}.
\]
With $p_j=2^{-j-10}$, $x=1+V/\chi$, $t_0=0$, and
$t_{j+1}=t_j+x+1+2xp_j^{-2}e^{64t_j}$, unioning the likelihood and
transferred-exposure exceptions of probability $p_j$ yields
$P_+\{W_B>\chi t_B\}<1/16$ uniformly in $B$. Correct selection forces
$t_B\ge c_\nu H$: otherwise $\Lambda_B\le\sqrt m$ with high probability,
and \eqref{eq:settings} together with $P_0$-selection at most $1/m$
contradicts the $5/8$ active selection rate. Rescaling gives
$128t_{j+1}\le e^{128t_j}$ for $j\ge1$ with $t_1=O(x)$; applying
$B-1$ clipped logarithms and $V=\Omega(\chi)$ proves
$C_B^*=\Omega(N\chi L_{B-1}(H))$, and hence
$T_B^*=\Omega(n_b\chi L_{B-1}(H))$ via $C_T\le bT$.

For a predetermined order, truncate its informative prefix at a large
constant multiple of its worst-instance expected cost (presampling pure
baseline reads into an independent noise tape so unboundedly many
zero-setting reads are covered). Under a uniform $s$-block support and
independent uniform best labels, conditional Fano requires
$\Omega(s\log m)$ nats, whereas comparing each row to its label-averaged
reference bounds mutual information by $O(s\sup\E C_T/(\chi Km))$.

\subsection{Capacity, small supports, and a full-batch upper bound}
Give each block $p$ slots, padding unused slots with baseline, and
assign them without replacement to $p$ layers. Within each layer,
partition the $K$ blocks into $J=\min\{K,\lceil C_JK/b\rceil\}$ balanced
groups. One pass uses $Jp$ legal rows of width at most $b$ and visits
each cell once. For any fixed cell set $F$ spanning at most $K/2$
blocks, its row neighborhood $\Gamma(F)$ after $d$ independent passes
satisfies
\begin{equation}\label{eq:main-forest}
 \Prob\{d|F|-|\Gamma(F)|\ge q\}
 \le \left(\frac{Cd|F|^2}{qJp}\right)^q .
\end{equation}
Joining same-row cells of $F$ by passwise spanning trees leaves at
least $K/2$ available block positions at each new tree edge; counting
compatible layer assignments and forest edges yields \eqref{eq:main-forest}.

To bound hits from an external cell set into $\Gamma(F)$ when $|F|=f$,
count deterministically every block containing more than $p/2$ cells of
$F$ (at most $2f/p$ blocks and $2f$ external cells). Every remaining
block retains at least $p/2$ unexposed layers, so sampling without
replacement bounds the joint hit probability of $r$ specified external
incidences by $(8f/(Jp))^r$ without requiring $f<m/2$, which extends
the residual analysis to $m<s$.

In batch one, a low-cost screen samples $\lfloor m/H^4\rfloor$ labels
per block and takes passwise medians over $\Theta(H)$ passes at
$\Theta(\chi H^2)$ baseline pairs per row. At scale
$t_0=\varepsilon/(sH)$ and cost $O(n_b\chi/H)$, it returns at most
$2s$ blocks $W$ such that outside $W$ at most $D_*=C_*sH^4$ cells have
$|\theta_{i,a}|>3t_0$, with failure probability $e^{-As}$. Hypergeometric
Chernoff bounds similarly cap the weak-cell load across any $O(ds)$ rows
at $O(dsH/\log(eH))$.

For $h\le s$, let $B_0\le B$ be the largest integer with
$\log^{(B_0-1)}H\ge2$ (or $B_0=B$ if smaller), and run $B_0-1$
compression batches followed by one joint validation batch. For
$j<B_0$, set
\[
 x_j=\log^{(B_0-j)}H,\quad
 d_j=\lceil C x_j\rceil,\quad l_j=\lceil me^{-4x_j}\rceil .
\]
Batch one pairs the screen with a full-slot table; stage $j\ge2$ uses
at most $2l_{j-1}$ slots per ordinary block, row threshold
$\varepsilon/(A_\tau s)$, and $\lceil C_Q\chi4^{j-1}\rceil$ pairs per row.
Stage $j$ retains labels with many high-magnitude rows (plus heavily
$W$-masked labels at $j=1$), and promotes up to $2\lceil s/2^j\rceil$
blocks exceeding $2l_j$ survivors to full known-block validation;
failure witnesses span at most $\lceil s/2^j\rceil$ blocks, so
\eqref{eq:main-forest} applies at arbitrary $m$.

Because each active block's best positive label is discarded at most
once, threshold losses sum over at most $s$ cells and stage noise errors
decay geometrically. Collisions leave at most $k=O(s/H)$ exceptional
discarded cells, which are recovered in the final batch on a fresh
certified layout alongside known-block Hadamard validation and shared-block
validation of the surviving lists (for $B=1$, the shared-block design
validates all $m$ slots directly). Since $x_{j+1}=e^{x_j}$,
\[
 \sum_{j=2}^{B_0-1} n_b\chi\,4^{j-1}/x_j^3=O(n_b\chi),
\]
slot rounding adds $O((K/b)\chi(\log H)^3)=o(n_b\chi)$, batch one costs
$O(n_b\chi x_1)$, and final validation costs $O(n_b\chi)$, with
$x_1=\Theta(L_{B-1}(H))$ on every execution path.

\subsection{Paying for higher confidence}
Let $t=\max(s,h)$ and $R=t/s$. Under the Gaussian zero instance with
stopped query matrix $A_0$, write $w(v)$ for the sum of within-block
ranges (including baseline) of a block-sparse vector $v$. Any single
assignment incurs combined regret exactly $cw(v)$ across $\pm cv$. For
$cw(v)>2\varepsilon$, binary data processing gives
\[
 \E_0\|A_0v\|_2^2\ge
 \frac{\sigma^2w(v)^2}{2\varepsilon^2}\kl(1/2,\delta).
\]
Averaging over $s$ random signed cells in distinct blocks yields the
$\Omega(N\chi h/s)$ setting lower bound in $U_B$.

For the matching upper bound in the intermediate range $1\le R<c_0 H$,
allow the screen to leave $O(tH^4)$ significant cells and give stage $j$
row precision $O(\chi 4^{j-1}(1+R/d_j))$, which costs
$O(\chi4^{j-1}(d_j+R))$ per pass table and caps collision excess at
$O(t/H)$ with failure $e^{-At}$. Local tail mass is bounded by
\[
 C\varepsilon\{\sqrt{R/H}+R/H+B_0/\log(eH)\},
\]
which stays below a small fraction of $\varepsilon$ for $c_0$ chosen
sufficiently small, while the fresh residual validator uses $O(\chi R)$
pairs per row. For $R\ge c_0H$, a single full-list shared-block batch
already costs $O(n_b\chi R)$, completing Theorem~\ref{thm:uniform}
without extra feedback batches.

\section{Polynomial computation and its boundary}\label{sec:computation}
\subsection{Which cells require a residual measurement}
After the screen, filter independent full-slot tables $j=0,\ldots,t$ in
parallel with degrees $d_j=d_0 2^j$, thresholds scaled by
$a_j=2^{3j/4}$, and row repetitions divided by $a_j^2$ along a shared
truncated-exponential precision schedule up to $H$. Because
$\sum d_j=O(\chi)$ and $\sum d_j/a_j^2=O(1)$, both the initial pass
count and total precision cost remain $O(n_b\chi)$.

Call a significant cell bad in table $j$ if at least $d_j/4$ of its rows
meet another significant cell. Passwise spanning forests on $r$ bad
cells contain at least $d_jr/8$ incident edges, bounding the probability
of $r$ or more bad cells by $e^{-c_\nu d_jHr}$. For $j\ge1$, positive optima first
isolated at degree $d_j$ were bad at degree $d_{j-1}$, so their count
is $O(g/d_j)$ and their omitted mass sums via $\sum a_j/d_j<\infty$.
At terminal degree $d$, at most $k=\lfloor A_0g/d\rfloor$ exceptional
discarded cells remain outside the directly validated blocks and
candidate lists.

\subsection{Certifying the fresh residual layout}
Two pre-observation statistics certify the fresh residual matrix in the
regimes of Theorem~\ref{thm:adaptive-compute}: the maximum $u$ of common
rows shared by any column pair, and the maximum $v$ of distinct third
columns meeting the first and second columns in different passes
(evaluated at finite budgets in Section~\ref{sec:experiments},
Fig.~\ref{fig:primary-formal}a). Single-pass pair collision has
probability $O(1/(Jm))$, and fixing two endpoints together with the
ordered pass pairs of $l$ third columns gives
\[
 \Prob\{v\ge l\}\le\frac{N^2}{l!}
       \left(\frac{C N d^2}{(Jm)^2}\right)^l.
\]
Bounded retries therefore yield bounded $u$, and bounded $v$ under
$g^{2/3}b^2/N\le s^{-\gamma}$, with failure probability at most
$\delta/64$.

Bounded $u$ alone suffices for an incident-median residual estimator
once $k$ is a small multiple of $d$, giving $g=O(\chi^2)$ at
$d=\Theta(\sqrt g)$. When $v$ is also bounded, matching disjoint pairs
within each row touched by $F$ and deduplicating edges leaves a simple
graph with at least $a_d(F)/(2u)$ edges and at most $v$ common neighbors
per vertex pair; counting length-two paths yields
\[
 a_d(F)\le u|F|(1+\sqrt{v|F|}),
\]
which certifies lossless expansion through $|F|=O(d^2)$ and hence covers
$2k=O(g/d)$ at $d=\Theta(g^{1/3})$.

Given expansion through $2k$, the residual decoder solves $\min\|z\|_1$
subject to bounding the sum of the $dk$ largest absolute row residuals
$\|y-Az\|_{[dk]}$. Fresh-layout occupancy restricts tail hits on any
$dk$ rows to $O(dk)$ incidences of normalized weight $O(k)$, matching
the filter's local tail bound and yielding $\ell_1$ recovery on the
exceptional set. Epigraph variables reduce the largest-$dk$ constraint
to a polynomial-size linear program with a small integer constraint
matrix, whose real-arithmetic complexity is independent of effect
magnitudes \cite{vavasis1996,dadush2024}. Outside the polynomial
certificate regimes, enumerating supports up to $2k$ accounts for the
sole exponential factor in \eqref{eq:runtime}.

Even at $\chi=1$ and $h=s$, instances with at most $C_-\chi$
unrestricted negative cells and all other cells bounded below by
$-M\varepsilon/s$ (fixed $C_-,M$, arbitrary positive effects) remain
polynomial-time solvable without a residual LP: choosing $d_0$ large
enough for $u\le u_0$ ensures that extreme negatives mask fewer than
$d/8$ rows per missed positive optimum, while largest-row occupancy
controls the moderate negatives.

\subsection{General-dimensional settings by ranked lists}
With $l_i$ candidate labels in block $i$, cyclic block-order assignment
packs one signed Hadamard cycle of order $P=2^{\lceil\log_2 b\rceil}$
into $\max(\max_i l_i,\lceil\sum_i l_i/b\rceil)$ groups, giving conditional
noise proxy $2\sigma^2\|w\|_2^2/(Pq)$ for any linear test $w$ at $q$
pairs per signed row. Retaining the top $\ell$ estimates per block bounds
any drop in a block's best retained true value by the true maximizer's
estimation error plus the average of the $\ell$ largest errors in that
block; these drops telescope across batches, and signed tests across at
most $s$ active blocks have squared norm at most $s/\ell$ for arbitrary
$K$.

Combining iterated-logarithm list shrinkage with working width
$\min\{b,\lceil\chi U_B^m\rceil\}$ proves Theorem~\ref{thm:rank-general}.
When $b\gg\chi U_B^m$, scalar observations can bottleneck $T$ even after
$C_T$ is optimal: at $K=b=s$, $m=2$, $\chi=1$, and fixed $\delta$, the
optimal labels carry $\Omega(s)$ nats whereas each conditional Gaussian
sum conveys at most $O(\log(es))$ nats, forcing
$T=\Omega(s/\log(es))$ despite $C_B^*/b=O(1)$.

\section{Finite-budget simulations}\label{sec:experiments}
ChatGPT (OpenAI) assisted with the mathematical derivations and with implementing the simulations. The supplement gives the derivations, instance generators, and evaluation rules. The accompanying archive provides executable code and the paired outcomes used below.

\begin{figure}[htbp]
\centering
\includegraphics[width=\textwidth]{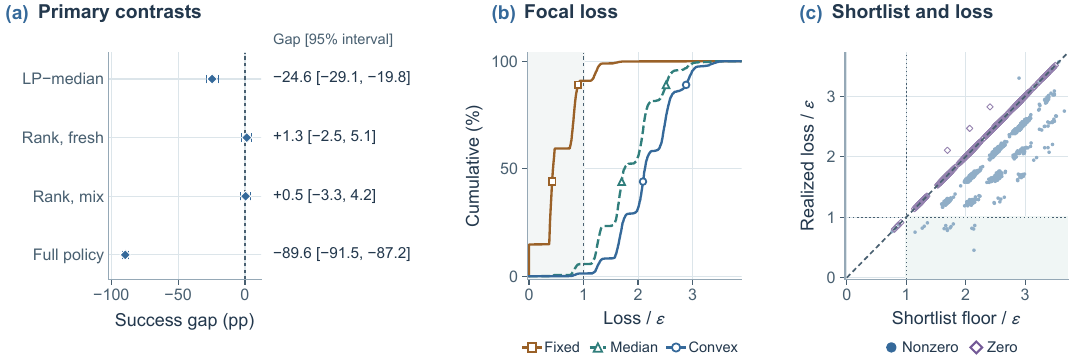}
\caption{Residual layout checks, full-policy comparison, and shortlist floor.
(a) Paired success-rate differences with simultaneous 95\% intervals
($2{,}048$ pairs per row). The rows compare LP versus incident median
on identical data at $s=k=6$, best-of-four $(u,v)$ ranking versus first
draw on fresh layouts and under pass-repetition risk, and two-batch
adaptive convex versus fixed orthogonal regression at $s=8$.
(b) Cumulative distributions of normalized loss at $T=1{,}072$ and
$C_T=2{,}584$. Success requires loss at most one. Median uses the first
layout and two batches.
(c) Adaptive convex loss against the best loss attainable from the
validated shortlist and baseline, both in units of $\varepsilon$. Zero
fits lie on the diagonal. Selecting a residual label can improve on the
shortlist floor.}\label{fig:primary-formal}
\end{figure}

Operational success requires normal termination with total assignment
loss $\mathcal R_\theta(\widehat x)\le\varepsilon$; scalar readings $T$,
settings $C_T$, and feedback batches are tracked separately. Disjoint
paired samples support four evaluations: an equal-budget feedback study,
a prespecified $342$-condition suite, a $256$-pair-per-stratum
matched-decoder study, and a $192$-pair-per-stratum coverage holdout
(Figs.~\ref{fig:primary-formal} and~\ref{fig:coverage}; protocols,
generators, and simultaneous intervals appear in the supplement).

\paragraph{Feedback at equal costs.}
At $s=m=b=64$ and $\chi=4$, with the same ridge-after-screening decoder
and identical budgets of $262{,}144$ readings and $3{,}178{,}496$
settings, four batches succeeded on $1{,}070$ of $1{,}200$ paired signed
instances, compared with $779$ for two batches---a gain of $24.25$
percentage points (paired 95\% interval $[20.23,28.11]$;
Fig.~\ref{fig:feedback}c). The two schedules differ only in whether row
rankings and the singleton shortlist are updated once more. Spending all
reading pairs on mixed rows without singletons succeeded on $1{,}199$ of
$1{,}200$ instances while using $31.96\%$ more settings. In a separate
orthogonal-group study across $s\in\{16,24\}$, both two-stage filtering
and fixed measurement succeeded on all $4{,}608$ paired instances at
matched final precision, with filtering using $55.81\%$ of the readings
and $55.73\%$ of the settings on average.

\paragraph{Layout checks and the convex residual.}
On identical residual data at $s=6$ with public cap $k=6$ and fresh
random layouts ($2{,}048$ paired instances per comparison), the convex
LP decoder succeeded on \MainOneA{} instances versus \MainOneB{} for the
incident median ($\MainOneDelta{}$ points, simultaneous 95\% interval
$[\MainOneLo{},\MainOneHi{}]$). Holding the convex decoder fixed,
selecting the best of four $(u,v)$-ranked layouts over the first draw
changed success by $+\MainTwoDelta{}$ points on fresh draws ($\MainTwoA{}$
vs.\ $\MainTwoB{}$) and $+\MainThreeDelta{}$ points under pass-repetition
risk ($\MainThreeA{}$ vs.\ $\MainThreeB{}$); both simultaneous intervals
cross zero (Fig.~\ref{fig:primary-formal}a). Comparing complete policies
at $s=8$ under identical budgets $T=1{,}072$ and $C_T=2{,}584$, the
two-batch adaptive convex policy succeeded on \MainFourA{} of $2{,}048$
paired instances (with all LP solves completing normally), compared with
\MainFourB{} for fixed signed Hadamard ridge regression
(Fig.~\ref{fig:primary-formal}b).

Whenever $\|y\|_{[r]}\le\eta$, the program $\min\|z\|_1$ subject to
$\|y-Az\|_{[r]}\le\eta$ returns $z=0$ uniquely. That occurred on
\EndpointZero{} of the $2{,}048$ instances, and on \EndpointFloor{} of
those \EndpointZero{} even an oracle restricted to the validated
shortlist and baseline exceeded $\varepsilon$
(Fig.~\ref{fig:primary-formal}c). Decomposing the mean adaptive convex
loss of $2.169\varepsilon$ relative to a four-cell residual oracle
assigns $0.636\varepsilon$ to the four-cell cap itself, $1.532\varepsilon$
to which extra labels the fit deployed, and $0.001\varepsilon$ to final
selection within the deployed set.

\begin{figure}[htbp]
\centering
\includegraphics[width=\textwidth]{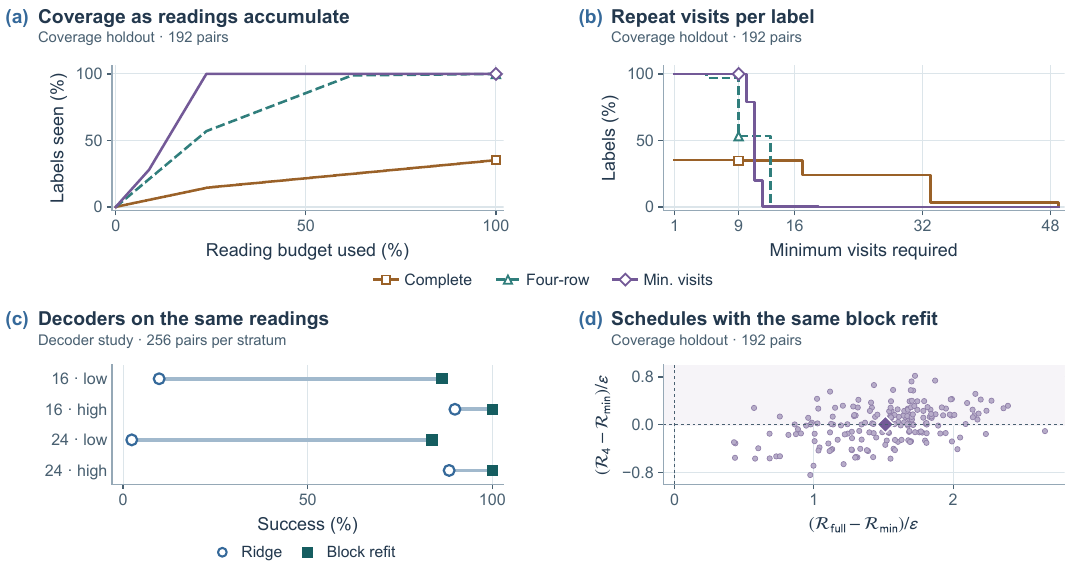}
\caption{Fixed-schedule label coverage and matched-data decoder accuracy.
(a,b) Coverage of $6{,}912$ labels at $s=24$, $T=8{,}256$ and $C_T=69{,}192$,
averaged over $192$ layouts per schedule. Panel (a) tracks cumulative
coverage. Panel (b) shows the share receiving at least the indicated
number of visits. The vertical line marks nine guaranteed visits.
(c) Success of two decoders on identical adaptive readings in a separate
holdout ($256$ pairs per stratum).
(d) Normalized loss differences against minimum visits on the same $192$
pairs as in (a,b), using a shared block refit. The diamond marks the paired
mean.}\label{fig:coverage}
\end{figure}

\paragraph{Isolating decoder and measurement effects.}
Across $s=16$ ($2{,}048$ labels) and $s=24$ ($6{,}912$ labels) at low
and high budgets ($256$ new paired instances per stratum), we first
evaluated acquisition plans under a shared global ridge decoder with
fixed penalty $0.1C_T/(Km)$. At the higher $s=24$ budget, the adaptive
plan succeeded on $226$ of $256$ instances versus $73$ for fixed
complete Hadamard cycles ($+59.77$ points, simultaneous 95\% interval
$[43.89,71.92]$); the other three same-ridge adaptive-versus-complete
intervals crossed zero. Holding the adaptive readings fixed and
switching the decoder from global ridge to two-stage block selection and
refitting raised low-budget successes out of $256$ from $25$ to $221$ at
$s=16$ and from $6$ to $214$ at $s=24$ (Fig.~\ref{fig:coverage}c).
Calibrating the residual radius on Gaussian noise eliminated all $422$
zero fits across the $1{,}024$ instances, yet both original and
calibrated convex residual policies succeeded on $0$ of $1{,}024$
instances.

\begin{table}[htbp]\centering
\caption{Independent signed coverage holdout: successes out of $192$ paired instances per stratum under the shared block-refit decoder. Lower budgets $(T,C_T)$ are $(3{,}904,20{,}528)$ at $s=16$ ($2{,}048$ labels) and $(8{,}256,69{,}192)$ at $s=24$ ($6{,}912$ labels); higher budgets double both counts.}\label{tab:visits-main}
\begin{tabular}{lrrrr}\toprule & \multicolumn{2}{c}{$s=16$} & \multicolumn{2}{c}{$s=24$}\\
Fixed schedule & Low & High & Low & High\\\midrule
Complete cycles & 12 & 182 & 0 & 54\\
Four-row cycles & 185 & 192 & 176 & 192\\
Minimum visits & 184 & 192 & 171 & 192\\
\bottomrule\end{tabular}\end{table}

In fixed acquisition, completing an entire Hadamard cycle on each group
before advancing visits at most $2{,}675$ of the $6{,}912$ labels at the
lower $s=24$ budget. The minimum-visit schedule distributes contrasts across
all groups and guarantees at least $9$ visits per label at both lower
budgets and $19$ at both higher budgets (Fig.~\ref{fig:coverage}a,b). In
an independent holdout of $192$ paired instances per stratum under the
same block-refit decoder (Table~\ref{tab:visits-main}), complete cycles,
four-row cycles, and the minimum-visit schedule succeeded on $0/192$,
$176/192$, and $171/192$ instances at the lower $s=24$ budget, and on
$12/192$, $185/192$, and $184/192$ at the lower $s=16$ budget.
At the lower $s=24$ budget, minimum visits reduced loss relative to complete
cycles on all $192$ pairs. Against four-row cycles, minimum visits gave
lower loss on $108$ pairs and higher loss on $84$ (Fig.~\ref{fig:coverage}d).
Across the four strata, all simultaneous 95\% intervals for the
minimum-visit minus four-row success difference included zero.

\section{Discussion}
In the benchmark, the value of another batch can be read off before any data are collected. If the iterated-log term exceeds both one and $h/s$, an extra opportunity to change the queries lowers the factor in both bounds. Once the confidence term dominates, further rounds leave the minimax orders unchanged. The predetermined-order result ties this gain to the use of earlier observations in choosing the next queries.

Efficient decoding introduces a separate question. At $m=b=s$ and $h=s$, the multiscale residual algorithm attains both adaptive orders in polynomial real-arithmetic time for $\chi=\Omega((s/\log(es))^{1/3})$. At $\chi=1$ its cost is within a factor $O((s/\log(es))^{1/3})$ of optimum. Whether a polynomial-time algorithm closes that gap for arbitrary signed effects remains open. For general dimensions, rank elimination already attains the optimal setting cost, even in regimes where wide queries create a separate bottleneck in reading complexity.

The simulations distinguish the cost of collecting a reading from the value of the labels it tests. In the lower-budget $s=24$ holdout, four-row cycles reached $99.96\%$ of the labels on average, and the minimum-visit design reached every label at least nine times. With a common decoder both schedules substantially improved on complete cycles. The paired intervals leave the ordering of the two high-coverage schedules unresolved.

\section*{Funding}
This research received no funding.

\section*{Declaration of competing interest}
The authors declare no competing interests.

\section*{Data and code availability}
The supplementary data and code archive contains the simulated outcomes, experiment code, and source data and scripts for the figures. Its README gives the reproduction instructions.

\section*{Declaration of generative AI and AI-assisted technologies in the manuscript preparation process}
The authors used ChatGPT (OpenAI) to assist with manuscript editing. Its use in mathematical development and experimental implementation is described in Section~\ref{sec:experiments}. The authors take responsibility for the content of the article.

\bibliographystyle{elsarticle-num}
\bibliography{references}
\end{document}

\raggedbottom
\begin{frontmatter}
\title{Supplementary material for\\[.3em]Feedback Rounds and Query Costs in Sparse Product Pure Exploration}
\author[sysu]{Yujiao Peng\fnref{equal}}
\author[ruc]{Qifan Yang\corref{cor}\fnref{equal}}
\ead{Y2771433453@ruc.edu.cn}
\address[sysu]{Sun Yat-sen University, China}
\address[ruc]{School of Business, Renmin University of China, Beijing, China}
\cortext[cor]{Corresponding author.}
\fntext[equal]{Yujiao Peng and Qifan Yang contributed equally to this work.}
\begin{abstract}
This supplement gives the lower bounds and measurement constructions for the batch--confidence tradeoff, the residual certificates for polynomial-time recovery, and the setting bounds for general dimensions. The experimental sections specify the simulation protocols, calibration and holdout splits, loss decompositions, and paired comparisons underlying the main article. The final appendix compares the reading bounds for an unrestricted product query.
\end{abstract}
\begin{keyword}
Proofs \sep Simulation protocols \sep Paired comparisons
\end{keyword}
\end{frontmatter}
\noindent\textbf{Guide to the supplement.}
Sections~\ref{sec:model} and~\ref{sec:main} repeat the model and main theorem statements for reference.
Sections~\ref{sec:lower}--\ref{sec:all-confidence} prove the information bounds,
and Sections~\ref{sec:budget-compute}--\ref{sec:rank-packing} develop the computational constructions and general-dimensional bounds.
Section~\ref{sec:finite-codes} treats finite measurement cycles.
Sections~\ref{sec:full-experiments}--\ref{sec:formal-studies} document the experiments.
\ref{sec:cho} compares unrestricted product queries.

\section{Model and performance criteria}\label{sec:model}
Block $i\in[K]$ has baseline label $0$ and $m$ nonbaseline labels. An
assignment $x$ with at most $b$ nonzero coordinates is a legal query.
One physical reading returns
\begin{equation}\label{eq:observation}
 Y=\mu_0+\sum_{i=1}^K\theta_{i,x_i}+\xi,\qquad \theta_{i,0}=0 .
\end{equation}
Both $\mu_0$ and the effects $\theta_{i,a}$ are unknown. The active block set
$S=\{i:\exists a,\ \theta_{i,a}\ne0\}$ has size at most $s$, and active
entries are arbitrary finite real numbers. The learner knows
$K,m,b,s,\varepsilon,\delta$ and a valid noise proxy $\sigma^2$. Write
$\mathcal F_{t-1}$ for the history of prior physical outcomes and private
randomness drawn before query $t$, including the current batch plan; each
fresh private draw is independent of $\mathcal F_{t-1}$. Conditional noise
satisfies
\begin{equation}\label{eq:mgf}
 \E[e^{\lambda\xi_t}\mid\mathcal F_{t-1},x_t]
       \le e^{\lambda^2\sigma^2/2},\qquad \lambda\in\mathbb R.
\end{equation}
Averaging $q$ fresh query--baseline differences cancels $\mu_0$, yields noise
proxy $2\sigma^2/q$, and costs $2q$ physical readings.

The learner stops almost surely and outputs a full assignment $\widehat x$
with no capacity limit and total assignment loss
\begin{equation}\label{eq:loss}
 \mathcal R_\theta(\widehat x)=
 \sum_{i=1}^K\{\max_{0\le a\le m}\theta_{i,a}-\theta_{i,\widehat x_i}\}.
\end{equation}
An $(\varepsilon,\delta)$-PAC algorithm satisfies
$\Prob_\theta\{\mathcal R_\theta(\widehat x)\le\varepsilon\}\ge 1-\delta$
for every admissible instance and noise law. We track both the physical
reading count $T$ and the cumulative setting count
\begin{equation}\label{eq:settings}
 C_T=\sum_{t=1}^T|\{i:x_{t,i}\ne0\}|\le bT .
\end{equation}
A baseline-only measurement contributes one reading and zero settings. A block tested on two successive readings contributes a setting on each. Under each protocol, the minimax reading and setting costs are the separate
infima over $(\varepsilon,\delta)$-PAC algorithms of their worst-instance
expected values of $T$ and $C_T$. A predetermined order fixes its query
sequence from private randomness before observing data, though its stopping
time may depend on feedback. In a strict $B$-batch algorithm, all queries and
the length of each batch are fixed before its first observation; empty
batches are allowed.

Let $N=Km$ be the number of nonbaseline cells and $n_b=N/b$ the reading-count scale for one full pass. Covering every cell requires at least $\lceil n_b\rceil$ physical readings. Fix $0<\nu\le1\le\zeta$. The signed
computational benchmark is
\begin{equation}\label{eq:benchmark}
 \begin{gathered}
 K=s^2,\quad 1\le b\le s,\quad s^\nu\le m\le s^\zeta,
 \quad H=\log(es),\\
 \log16\le h=\log(1/\delta)\le s,\qquad
 \chi=\sigma^2s^2/\varepsilon^2\ge1 .
 \end{gathered}
\end{equation}
Theorem~\ref{thm:uniform} extends \eqref{eq:benchmark} to all $h\ge\log16$,
and Theorem~\ref{thm:rank-general} treats arbitrary $(K,m,b,s)$. Natural
logarithms are used unless noted otherwise. Implied constants may depend on
$\nu,\zeta$, and asymptotic bounds hold for $s\ge s_0(\nu,\zeta)$ uniformly
in the remaining parameters; below $s_0$, averaging
$\lceil16\chi\log(2N/\delta)\rceil$ single-cell pairs per cell ensures
simultaneous $\varepsilon/(2s)$ estimation error and hence regret at most
$\varepsilon$. Running-time bounds count real arithmetic operations,
comparisons, square roots (as in the cited LP solver), and index operations;
bit complexity additionally depends on input encoding lengths.

\section{Main results}\label{sec:main}
For $u\ge1$ set $L_0(u)=u$ and $L_{q+1}(u)=\max\{1,\log L_q(u)\}$.
The clipped logarithm records the point at which another round ceases
to lower the order of the query cost. We use $\log^*u$ for the number
of natural logarithms needed to reach at most one.

\begin{theorem}[Cost of confidence and feedback]\label{thm:uniform}
Retain the dimension and noise-scale conditions in \eqref{eq:benchmark}
and allow every $h\ge\log16$. Set $U_B=\max\{L_{B-1}(H),h/s\}$.
For sufficiently large $s$, the two strict $B$-batch minimax costs are
\[
  T_B^*=\Theta(n_b\chi U_B),\qquad C_B^*=\Theta(N\chi U_B).
\]
Constants and the starting scale depend only on $\nu,\zeta$.
One algorithm attains both upper bounds on every execution path.
The lower bounds hold for independent Gaussian noise with known
zero intercept.
\end{theorem}

\begin{corollary}[Endpoint protocols]\label{thm:extremes}
A predetermined order with feedback-dependent stopping pays the same
orders as one batch. Unrestricted adaptation pays the common normalized
factor $\chi R$, where $R=\max\{1,h/s\}$. For $z=\max\{2,h/s\}$ let
$j_z(H)=\min\{j\ge0:L_j(H)\le z\}$. A uniform cap of $1+j_z(H)$
batches suffices at the adaptive rate. With either adaptive cost held
within a fixed multiplier $L$, a cap of at least
$1+j_z(H)-O_{\nu,\zeta,L}(1)$ is necessary.
\end{corollary}

\begin{theorem}[Optimization at the adaptive rate]\label{thm:adaptive-compute}
Under the full benchmark \eqref{eq:benchmark}, including $h\le s$,
write $g=1+h/H$. Both optimal pathwise adaptive cost orders are
attainable in $\log^*H+O(1)$ batches in polynomial real-arithmetic
time if any of the following conditions holds:
\begin{enumerate}[label=\arabic*)]
 \item $g=O(\chi^2)$;
 \item $g=O(\chi^3)$ and $g^{2/3}b^2/N\le s^{-\gamma}$ for fixed
       $\gamma>0$;
 \item $b=O(\chi)$.
\end{enumerate}
The constants may depend on $\nu,\zeta,\gamma$ and the fixed implied
multipliers. In particular, at $m=b=s$ and $h=s$, the condition
$\chi=\Omega((s/H)^{1/3})$ suffices for unrestricted signed effects.
For every fixed $\eta>0$, the whole benchmark admits the same cost
and batch orders in
\begin{equation}\label{eq:runtime}
 \operatorname{poly}_{\nu,\zeta,\eta}(s,m,T_{\max})e^{\eta h/\chi}
\end{equation}
operations, with query constants allowed to depend on $\eta$.
Here $T_{\max}$ is the deterministic observation cap.
\end{theorem}

\begin{theorem}[Setting cost in general dimensions]\label{thm:rank-general}
For arbitrary $1\le s,b\le K$, $m\ge2$, $\chi\ge1$ and
$h\ge\log16$, put $H_m=\log(em)$ and
$U^m_B=\max\{L_{B-1}(H_m),h/s\}$. Then
\[
 C_B^*=\Theta(Km\chi U^m_B),\qquad
 T_B^*=\Omega((Km/b)\chi U^m_B).
\]
A polynomial-time algorithm with at most $B$ batches attains the
setting upper bound pathwise with $T=O((Km/b)(b+\chi U^m_B))$.
Thus it attains both optimum orders when $b=O(\chi U^m_B)$ with a
fixed multiplier. The bounds before this last condition use
absolute constants.
\end{theorem}

\section{Lower bounds with actual stopping}\label{sec:lower}
Throughout this section observations are independent Gaussian with
variance $\sigma^2$ and known intercept $\mu_0=0$ (fixing $\mu_0=0$ only
strengthens the lower bounds). All algorithm random seeds are independent
of the instance and of observation noise.

\subsection{An absent-group reference}
Partition the $K=s^2$ blocks into $s$ groups of $s$ blocks. Independently activate each group with probability $1/2$. An active group has one uniformly selected cell $e=(i,a)$ among its $N_g=sm$ cells, with effect $\Delta=8\varepsilon/s$; all other effects are zero. The prior is supported on admissible instances. A PAC output misses at most $s/8$ active cells on its success event. Thus, for a uniformly sampled group conditioned to be active, its average probability of selecting its true cell is at least
\begin{equation}\label{eq:active-success}
 1-2(1/8+\delta)\ge5/8.
\end{equation}
The output selects at most $s$ cells in that group, one per physical block.

For the analysis, sample a group $g$, its other-group parameters, a uniform cell $e$, and the algorithm seed. Let $P_+$ activate $g$ at $e$ and let $P_0$ leave $g$ empty, with the same other parameters. The side variables are not revealed to the algorithm. Under $P_0$, $e$ is independent of the transcript; hence
\begin{equation}\label{eq:null-list}
 P_0\{e\text{ is selected by the output}\}\le1/m.
\end{equation}

\begin{lemma}[Reference budget and stopped information]\label{lem:reference}
Let $c=\sup_\theta\E_\theta\cost$ and $V=c/(Km)$. Let $w_j(e)$ be the number of settings of $e$ in batch $j$. For every $u>0$,
\begin{equation}\label{eq:absent-budget}
 P_0\{w_j(e)\ge u\}\le 2V/u.
\end{equation}
For total exposure $w(e)$, $\E_0 w(e)\le2V$ and
\begin{equation}\label{eq:reverse-kl}
 \KL(P_0\Vert P_+)=\frac{\Delta^2}{2\sigma^2}\E_0w(e)
 \le 64V/\chi.
\end{equation}
The equality is for the joint experiment including the side variables.
\end{lemma}
\begin{proof}
Let $C_{j,g}$ denote settings in group $g$. Summing empty-group budgets before averaging over $g$ gives
\[
 \frac1s\sum_g\E[C_{j,g}\mid g\text{ empty}]
 =\frac2s\sum_g\E[\one_{\{g\text{ empty}\}}C_{j,g}]
 \le2c/s.
\]
Uniform averaging over $N_g$ cells divides this by $N_g$, giving $2c/(sN_g)=2V$. The same identity applies to total cost, and Markov's inequality proves \eqref{eq:absent-budget}.

Given the past, the two Gaussian means differ by $\Delta$ exactly when cell $e$ is queried. The likelihood chain rule therefore gives the stopped KL identity. One may first stop after a deterministic number of physical observations and then pass to the limit. Almost-sure termination and finite expected informative exposure justify the passage; an infinite expected cost already satisfies the lower bound. Baseline observations carry zero likelihood increment. Finally $\Delta^2/\sigma^2=64/\chi$.
\end{proof}

\begin{proposition}[Statistical coverage]\label{prop:coverage}
Every PAC algorithm under \eqref{eq:benchmark} has
\[
 \sup_\theta\E_\theta\cost\ge c_0Km\chi,
 \qquad \sup_\theta\E_\theta T\ge c_0n_b\chi.
\]
Independently of noise level, the corresponding coverage lower bounds are $c_1Km$ and $c_1 n_b$ for $\delta\le1/16$.
\end{proposition}
\begin{proof}
Pinsker's inequality, \eqref{eq:null-list}, and \eqref{eq:reverse-kl} bound the active success probability by $1/m+\sqrt{32V/\chi}$. Comparing with \eqref{eq:active-success} gives $V\ge c_0\chi$ for sufficiently large $m$. The capacity inequality transfers the bound to $T$.

For coverage, fix a cell and compare the two instances assigning it effects $2\varepsilon$ and $-2\varepsilon$. Their successful output sets are disjoint. On a zero-instance path that never sets the cell, both alternatives have the same transcript, even under adaptive stopping. Thus the zero-instance probability of never setting that cell is at most $2\delta$. Summing over all $Km$ cells bounds the expected number of distinct queried cells by $(1-2\delta)Km$ from below. Repetitions only increase settings.
\end{proof}

\subsection{Predetermined sequences may stop but cannot redirect}
\begin{proposition}[Order lower bound]\label{prop:order-lower}
A predetermined-order PAC algorithm satisfies
\[
 \sup_\theta\E_\theta\cost=\Omega(Km\chi H),\qquad
 \sup_\theta\E_\theta T=\Omega(n_b\chi H).
\]
\end{proposition}
\begin{proof}
Choose a uniform $s$-block support $U$ and independent uniform best labels $Z_i\in[m]$ on $U$, each with effect $\Delta=8\varepsilon/s$. Give $U$ to the analysis decoder, not to the query planner. Let $c$ be the worst-instance expected setting cost and truncate after a virtual setting budget $Ac$, for a sufficiently large numerical $A$. Conditional on the seed, the virtual query prefix is independent of $U$ and $Z$.

A predetermined sequence may contain arbitrarily many zero-setting actions. With $\mu_0=0$, all such responses are pure independent Gaussian noise and can be presampled into an additional independent random tape. Remove these actions from the informative sequence and simulate their stopping effects using that tape. Every remaining action costs at least one setting, so the prefix before the next action would exceed $Ac$ is finite. If the informative sequence ends, its remaining noise-only decisions are still measurable from the tape. The truncated output differs from the original only when its setting cost exceeds $Ac$, an event of probability at most $1/A$ for each instance.

On the joint success and no-truncation event at most $s/8$ labels are wrong.
The average per-block error $\bar p$ is at most $1/8+\delta+1/A<1/3$
for fixed $A$ large enough. Replace a baseline answer by any label for
this information bound. Conditional entropy subadditivity and the
single-coordinate Fano bound give
$H(Z\mid Y,U,\omega)\le s[h_2(\bar p)+\bar p\log(m-1)]$,
where $h_2$ is binary entropy in nats. Since the prior entropy is
$s\log m$, it follows that
\begin{equation}\label{eq:fano}
 I(Z;Y\mid U,\omega)\ge c_2s\log m,
\end{equation}
where $Y$ is the complete virtual prefix and $\omega$ includes the auxiliary tape. For each $U$, use as a reference the instance assigning effect $\Delta/m$ to every nonbaseline label of each block in $U$. For a predetermined row with $d_U$ selected blocks in $U$, the squared difference from this reference, averaged over $Z$, is at most $\Delta^2d_U/m$. Averaging over $U$ yields $\E d_U=(s/K)d$, where $d$ is row width. The reference-distribution upper bound on mutual information therefore yields
\[
 I(Z;Y\mid U,\omega)
 \le \frac{\Delta^2}{2\sigma^2m}\frac{s}{K}Ac
 =O\left(\frac{s c}{\chi Km}\right).
\]
Comparison with \eqref{eq:fano} proves the setting bound, since $\nu\log s\le\log m\le\zeta\log s$. For the query bound, truncate at $A\sup_\theta\E_\theta T$ physical queries instead; their total width is at most $b$ times their count, and the same information bound applies.
\end{proof}

\subsection{Exposure grows by at most one exponential layer per batch}
Write $W_j(e)=\sum_{k\le j}w_k(e)$ and let $\Lambda_j$ be the likelihood ratio $dP_+/dP_0$ up to the end of batch $j$, including the side variables. Under $P_+$,
\begin{equation}\label{eq:lr}
 \log\Lambda_j=I_j/2+M_j,\qquad
 I_j=(64/\chi)W_j(e),\qquad \langle M\rangle_j=I_j.
\end{equation}
Applying Markov's inequality to $\exp(M_j-I_j/2)$ (with localization for unbounded finite batch lengths) gives, for $A_0\ge64$ and all $t,u>0$,
\begin{equation}\label{eq:lr-tail}
 P_+\{W_j\le\chi t,\ \log\Lambda_j>A_0t+u\}\le e^{-u}.
\end{equation}

Because $\{w_{j+1}\ge\ell\}$ is measurable before batch $j+1$ is read, \eqref{eq:absent-budget} transfers it from $P_0$ to $P_+$ at multiplicative cost at most $\exp(A_0t+u)$ on the low-likelihood event at batch $j$, where the strict batch rule enters.

\begin{lemma}[Exposure needed for correct selection]\label{lem:need-exposure}
There is a small $c_3=c_3(\nu)>0$ such that a PAC algorithm cannot satisfy $P_+\{W_B>c_3\chi H\}\le1/16$ for all sufficiently large $s$.
\end{lemma}
\begin{proof}
By \eqref{eq:lr-tail}, on $W_B\le c_3\chi H$ the probability of $\Lambda_B>\sqrt m$ is at most $e^{-c_4H}$ if $c_3$ is sufficiently small in terms of $\nu$. On the complementary likelihood event, \eqref{eq:null-list} gives active success probability at most $m^{-1/2}$. Adding the two exceptional probabilities contradicts \eqref{eq:active-success}.
\end{proof}

\section{Legal layouts and signed stability}\label{sec:geometry}
Throughout, constants may depend on the fixed exponents $\nu,\zeta$ and may be increased to make a stated failure probability at most $e^{-As}$ for any prescribed fixed $A$. Every row across all passes of a single-batch design is fixed before that batch is read.

\subsection{Balanced layouts}
Give each block $p$ slots, padding unused slots with baseline. In one pass, independently permute each block's slots into $p$ layers. In each layer, independently permute the $K$ blocks into $J=\min\{K,\lceil C_JK/b\rceil\}$ predetermined groups of sizes differing by at most one. There are $P=Jp$ rows per pass. A row contains at most one slot per block and at most $b$ blocks since $\lceil K/J\rceil\le b$. A column is a block-slot pair, or \emph{cell}; each column occurs once per pass. Query a row and a fresh all-baseline action to obtain its sum. For $D$ passes, let $A$ be the resulting zero-one incidence matrix and let $\Gamma(F)$ denote the rows touching a cell set $F$.

\begin{lemma}[Forest and load bounds]\label{lem:forest}
For a fixed cell set $F$ with $f=|F|$ and at most $K/2$ distinct physical blocks represented in $F$ and collision excess $a_D(F)=Df-|\Gamma(F)|$,
\begin{equation}\label{eq:forest}
 \Prob\{a_D(F)\ge q\}\le
 \left(\frac{C D f^2}{qJp}\right)^q,\qquad q\ge1.
\end{equation}
For a fixed active block set of size at most $s$, with probability at least $1-e^{-As}$, the total active-slot load in any set of at most $cDs$ rows is at most $C_{A,c,\zeta}DsH/\log(eH)$. This holds for $1\le p\le m\le s^\zeta$, $1\le D\le C'H$, and also for any fixed larger multiple of $s$ in place of $c s$.
\end{lemma}
\begin{proof}
In each row occupied by $F$, join its cells by a tree. Excess at least $q$ supplies $q$ edges from the union of these passwise forests. A specified forest with $v$ endpoints and $q_r$ edges in one pass has at most $p^{v-q_r}$ compatible layer assignments. The denominators from sampling the layer positions without replacement are at least $(p/e)^v$. Two slots from one block cannot occupy the same layer. Conditional on compatible layers, reveal the group slots along the trees. Each new endpoint must match its predecessor's group with probability at most $C/J$, because in any layer the exposed endpoints represent distinct blocks, of which there are at most $K/2$ (even when $f>K/2$). Since $v\le2q_r$, the probability of the forest is at most $(C/(Jp))^{q_r}$. There are at most $\binom{D\binom{f}{2}}{q}$ choices of edges. Multiplying over independent passes and using $\binom{n}{q}\le(en/q)^q$ gives \eqref{eq:forest}.

For the load bound, count all $p$ slots of each active block, including any padded ones, which can only increase the load. In a fixed layer the active blocks occupy a uniform sample of group slots. Its count in a specified row union is hypergeometric, with the usual binomial moment-generating upper bound. Layers and passes have independent group permutations. A fixed collection of at most $cDs$ rows therefore has a Poisson-type moment bound with mean $\mu=O_c(Ds)$. The number of row collections has logarithm at most
$O_{c,\zeta}(DsH)$ since $p\le s^\zeta$. Put
$t=C_{A,c,\zeta}DsH/\log(eH)$. For sufficiently large $s$,
$\log(t/(e\mu))\ge\tfrac12\log(eH)$ after increasing its constant.
The Chernoff exponent $t\log(t/(e\mu))$ therefore pays for that entropy
and $As$. Similarly, if $\log(DJp)=O(H)$, a union over at most $c s$ rows
has entropy $O(sH)$ and any fixed such collection has mean active
load $O(s)$, giving $O_{A,c}(sH/\log(eH))$
simultaneously on all such collections with failure $e^{-As}$
even when $D>H$.
\end{proof}

\paragraph{Conditional occupancy.}
An incidence specifies a cell and a pass. Given the layer maps and some exposed group positions, the positions of other specified blocks in a layer are sampled without replacement from the free positions. If $r$ such blocks must all land in a union of groups with $a$ free positions out of $n$ free positions in the layer, their conditional probability is $(a)_r/(n)_r\le(a/n)^r$, where $(a)_r$ is the falling factorial. With at most $K/2$ positions exposed and $g$ target groups, $a/n\le4g/J$.

First fix a block set $U$ of size at most $2s$ and expose its group positions in every layer. Given the layer maps and these positions, any $q$ specified distinct incidences of cells outside $U$ all share a row with $U$ with probability at most
\begin{equation}\label{eq:block-hits}
 (4|U|/J)^q.
\end{equation}
In each layer the target union consists of at most $|U|$ groups. The remaining block positions are uniform, with independent permutations across layers and passes.

For the residual calculation, fix disjoint cell sets $F$ and $L$ with $f=|F|=o(K)$.
Call a block heavy if it contains more than $p/2$ cells of $F$. At most $2f/p$ blocks are heavy, and together they contain at most $2f$ distinct cells of $L$. Count those cells deterministically and apply the probability estimate below only to the remaining light-block cells $L_{\rm light}$; this partition applies for every $f$, including $f\ge p/2$. Expose the exact layer and group position of each cell of $F$ in each pass. Once these layers are specified, the exposure restricts the remaining layer maps and the group permutations separately. A block with $f_i$ exposed cells assigns its other labels uniformly without replacement to its $p-f_i$ remaining layers, independently across blocks. In a pass, let $g_\lambda$ count the groups touched by $F$ in layer $\lambda$, so $\sum_\lambda g_\lambda\le f$. A cell in $L$ cannot use a position already exposed for its block, because its layer differs from those of the exposed labels of that block.

For specified incidences of $L_{\rm light}$, first apply the group bound given their layer choices. For $r_i$ specified cells in one block in one pass, average the resulting product of the weights $4g_\lambda/J$ over the $r_i$ distinct layers assigned to them. The elementary symmetric mean of order $r_i$ is at most the arithmetic mean to the power $r_i$: averaging two nonnegative weights with fixed sum preserves terms using neither or one of them and increases terms using both, so repeatedly averaging the largest and smallest weights yields the constant vector. Since $p-f_i\ge p/2$, the averaged product is at most $(8f/(Jp))^{r_i}$. Multiplying over blocks and passes bounds the probability that any $q$ specified incidences all hit $\Gamma(F)$. Thus for $Z$ counting such hits from $L_{\rm light}$ under the unconditioned layout law,
\begin{equation}\label{eq:factorial-hits}
 \begin{gathered}
 \E\left[\binom Zq\,\middle|\,F\text{ slots}\right]
 \le \binom{D|L_{\rm light}|}{q}\left(\frac{Cf}{Jp}\right)^q
 \le\frac{\mu_f^q}{q!},\\
 \mu_f=\frac{CD|L_{\rm light}|f}{Jp},
 \end{gathered}
\end{equation}
so for $q\ge1$ the tail is at most $(e\mu_f/q)^q$ uniformly over the exposed positions.

\begin{lemma}[Certified expansion and prefix stability]\label{lem:expansion}
Fix $\alpha\in(0,1/8)$. With a sufficiently large constant number $D$ of passes and $p=m$, a balanced layout satisfies
\begin{equation}\label{eq:expansion}
 |\Gamma(F)|\ge(1-\alpha)D|F|\qquad(1\le |F|\le L)
\end{equation}
with probability at least $1/2$, whenever $L\le C_LsH^{4}$ with fixed $C_L$. This property can be checked by finite enumeration. With $p=1$, $J=\min\{K,\lceil C_JK/b\rceil\}$, and $D=\lceil C_DH\rceil$, the same conclusion holds for all block sets of size at most $2s$.

For any vector $z$ whose support and every subset of it satisfy \eqref{eq:expansion},
\begin{equation}\label{eq:rip1}
 \|Az\|_1\ge(1-2\alpha)D\|z\|_1.
\end{equation}
If rows are deleted so that each supported column loses at most $\kappa D$ incidences, the right side becomes $(1-2\alpha-\kappa)D\|z\|_1$.
\end{lemma}
\begin{proof}
For the cell assertion, union bound \eqref{eq:forest} with $q=\lceil\alpha Df\rceil$ over $\binom{Km}{f}$ sets. For $2\le f\le L$, $\log(Jm/f)\ge \nu\log s-O(\log H)$ and $\log(eKm/f)=O_\zeta(H)$. Taking $D$ large makes the sum at most $1/2$; singletons have no excess. For the block assertion, use $p=1$, $f\le2s$, and $J=\min\{K,\lceil C_JK/b\rceil\}$; taking $C_J$ large and then $D=\lceil C_DH\rceil$ with $C_D$ large absorbs $\log(eK/f)$ uniformly.

Order the nonzero coordinates by decreasing absolute value and in every row designate its earliest coordinate as first. For each prefix of $r$ coordinates, expansion leaves at most $\alpha Dr$ nonfirst incidences. Summation by parts bounds their weighted mass by $\alpha D\|z\|_1$, and the rowwise triangle inequality gives \eqref{eq:rip1}. Deleted rows remove at most $\kappa D\|z\|_1$ additional mass.
\end{proof}

When randomness relative to the true instance is needed, draw and certify a layout at most $L_{\rm try}=\lceil C s\rceil$ times before taking observations (adding no feedback batch), returning the baseline assignment if all attempts fail (probability at most $2^{-L_{\rm try}}$). Conditional on preceding history and acceptance, the first accepted layout follows the original law conditioned on \eqref{eq:expansion}, inflating any geometric failure bound under the unconditioned law by at most two; \eqref{eq:mgf} then applies conditional on the accepted plan.

\subsection{A compressed first-batch screen}
\begin{lemma}[Screening dense active blocks]\label{lem:screen}
Set $t_0=\varepsilon/(sH)$. In one predetermined batch of $o(n_b\chi)$ queries, an algorithm produces a set $W$ with a hard cap $|W|\le2s$. Except with probability $e^{-As}$, $W$ has at most $k_0=\lceil C_As/H\rceil$ blocks outside $S$ and
\begin{equation}\label{eq:screen}
 S_1=\{(i,a):i\notin W,\ |\theta_{i,a}|>3t_0\},
 \qquad |S_1|\le D_*:=C_AsH^{4}.
\end{equation}
For a fixed instance, the family of possible successful $W$ has logarithmic cardinality $O_A(s)$.
\end{lemma}
\begin{proof}
Each block independently samples $\ell=\lfloor m/H^{4}\rfloor$ labels without replacement ($\ell\ge m/(2H^{4})$ for large $s$). Reindex the sampled labels into $\ell$ slots and use an odd number $R=\Theta(H)$ of independent balanced passes, averaging $q=\lceil C_q\chi H^2\rceil$ paired readings per row. Estimate each sampled label by its passwise median, declare a block discovered at the batch boundary if one estimate has magnitude above $2t_0$, and return baseline if $|W|>2s$.

Fix the sampled labels. Call a cell bad if its estimate differs from its true effect by more than $t_0$. In a row containing no other active block, the row mean equals that cell's signed effect; hence if all cells of a fixed $k_0$-cell set $F$ are bad, at least $M/2$ of its $M=Rk_0$ incidences either meet another active block or have absolute row noise above $t_0$.

If at least $M/4$ incidences meet another active block, choose an active partner for each. Fair independent red--blue coloring of $F$'s blocks and $S$ leaves at least $M/16$ incidences in expectation with a red source and a blue partner, since source and partner lie in distinct blocks. Conditioning on blue block positions and applying \eqref{eq:block-hits}, a union over colorings and incidence subsets bounds this branch by
\[
 2^{s+k_0+M}(4s/J)^{M/16},
\]
allowing repeated partners and multiple sampled labels per block.

Otherwise at least $M/4$ incidences have large noise. Since $J\ell/k_0\ge c m/H^3\ge s^{\nu-o(1)}$, \eqref{eq:forest} gives $\Prob\{a_R(F)>M/16\}\le e^{-cMH}$; on its complement, at least $3M/16$ distinct rows of $\Gamma(F)$ have noise above $t_0$. Conditioning on the complete batch plan and unioning over row subsets and signs bounds this branch by
\[
 4^M\exp\!\left(-\frac{3Mqt_0^2}{64\sigma^2}\right),
\]
even when another measurement table shares the batch.

There are at most $\binom{K\ell}{k_0}=\exp(O(k_0H))$ witnesses $F$. Choosing $C_J$ to absorb $2^M$ and then choosing $R=\Theta(H)$ and $C_q$ ($qt_0^2/\sigma^2\ge C_q$) ensures fewer than $k_0$ bad cells with probability $1-e^{-2As}$.

Every falsely declared block and every sampled label with $|\theta_{i,a}|>3t_0$ outside $W$ requires a bad cell. Averaging over label draws for any fixed subset of $S$, let $d_i$ count labels with $|\theta_{i,a}|>3t_0$ in block $i$ and $N_i$ count those sampled. Applying the symmetric-mean inequality to weights $e^{-1}$ on those $d_i$ labels and $1$ elsewhere gives
\[
 \E e^{-\sum_i N_i}\le\exp\!\left(-(1-e^{-1})\,\frac{\ell}{m}\sum_i d_i\right).
\]
If $\sum_i d_i>D_*$, Markov's inequality at $\sum_i N_i\le k_0$ and a union over at most $2^s$ subsets of $S$ prove \eqref{eq:screen}. On the combined success event $|W|\le s+k_0\le2s$, and every successful $W$ lies in
\[
 \mathcal W_\theta
 =\{U\cup V: U\subseteq S,\ V\subseteq [K]\setminus S,\ |V|\le k_0\},
\]
which has size at most $2^s\sum_{r\le k_0}\binom Kr=\exp(O_A(s))$. The pathwise physical query count is $2RJ\ell q=O(n_b\chi/H)$.
\end{proof}

\subsection{Noise on predictably chosen rows}
\begin{lemma}[Joint conditional noise bounds]\label{lem:noise}
Suppose the rows measured at level $j$ are chosen from past levels, and each selected row uses $q_j$ fresh pairs. Write $\xi_{j,r}=0$ for an unmeasured row. Let $\mathcal P_j$ be the history at the start of the level, including its layout and plan. For a row set $U$ determined by $\mathcal P_j$,
\begin{equation}\label{eq:l1noise}
 \Prob\left\{\sum_{r\in U}|\xi_{j,r}|>
 C\sigma\sqrt{|U|\,(u+|U|)/q_j}\,\middle|\,\mathcal P_j\right\}\le e^{-u}.
\end{equation}
For a finite family determined by $\mathcal P_j$, the simultaneous bound uses the logarithm of the family size added to $u$.
\end{lemma}
\begin{proof}
For fixed signs, disjoint row pairs have squared physical coefficients summing to at most $2|U|/q_j$. Sequential conditioning in \eqref{eq:mgf} gives proxy $2\sigma^2|U|/q_j$; a scalar Chernoff bound and unions over $2^{|U|}$ signs and the prescribed family complete the proof.
\end{proof}

When $p\ge m/H^c$, $D\le C\log H$, and a fixed significant set $G$ has size at most $D_*$, \eqref{eq:forest} gives
\begin{equation}\label{eq:small-collisions}
 a_D(G)\le C_As/H
\end{equation}
with probability $1-e^{-As}$ (and $a_D(G)\le C_As$ with failure $e^{-\Omega(sH)}$ for constant $D$ and $p=m$). Rows of multiplicity $r\ge2$ contribute $r\le2(r-1)$ incidences, so at most $2a_D(G)$ incidences of $G$ collide. After a separate screen, $G=S_1$ is fixed by history; when screening and the first mixed table share batch one, union the fixed-set bound over $W\in\mathcal W_\theta$ satisfying \eqref{eq:screen}.

\section{Upper bounds and explicit budget caps}\label{sec:upper}
\subsection{A fixed design with a shared block layout}
The same fixed design validates both predetermined-order full lists and final surviving shortlists.

\begin{lemma}[Fixed validation of blockwise lists]\label{lem:fixed-validator}
Suppose each block has at most $p\le m$ requested labels, with baseline padding allowed. Under \eqref{eq:benchmark}, a predetermined design estimates an array $z$ supported on at most $s$ blocks such that
\[
 \mix{z-\theta^{\rm list}}\le\varepsilon/2,
 \qquad \mix{v}:=\sum_i\max_a|v_{i,a}|,
\]
with failure probability at most $e^{-As}$. The design uses
$O_A((K/b)p\chi H)$ observations and $O_A(Kp\chi H)$ settings
on every path. The target contains the requested labels, with zero
values fixed on all padded slots.
\end{lemma}
\begin{proof}
Use the block version of Lemma~\ref{lem:expansion} with $D=\lceil C_DH\rceil$ passes and $J=\min\{K,\lceil C_JK/b\rceil\}$ groups. Fix one certified block incidence matrix $E$ and reuse it across all $p$ label slots, repeating each group--baseline pair $q=\lceil C_q\chi\rceil$ times.

For any block set $U$ with $|U|\le2s$, $E_Uv$ has at most $2Ds$ nonzeros, so \eqref{eq:rip1} gives $\|E_Uv\|_2^2\ge c(D/s)\|v\|_1^2$. Thus $E_U^\top E_U$ is invertible. Writing $G_U=(E_U^\top E_U)^{-1}$, duality gives
\begin{equation}\label{eq:gram-dual}
 w^\top G_Uw\le C s/D\quad\text{whenever }\|w\|_\infty\le1.
\end{equation}
Form the least-squares array $\widetilde\theta_U$ for every $U$. On any $U\supseteq S$, decomposing a mixed-norm test into slot vectors $w_a$ with disjoint block supports and introducing independent Rademacher signs $\epsilon_a$ gives
\begin{equation}\label{eq:shared-gram}
 \sum_a w_a^\top G_Uw_a
 =\E_\epsilon\left[\left(\sum_a\epsilon_aw_a\right)^\top
 G_U\left(\sum_a\epsilon_aw_a\right)\right]
 \le Cs/D,
\end{equation}
since $\|\sum_a\epsilon_a w_a\|_\infty\le1$; reusing $E$ across slots thus avoids an extra factor $p$ from separate per-slot bounds.

By sequential conditional MGFs, each test noise has proxy at most $C\sigma^2s/(qD)$. Unioning over at most $\sum_{r\le2s}\binom Kr(2p)^r=\exp(O(sH))$ support--test pairs with $qD\ge C_A\chi H$ bounds all correctly specified errors by $\mix{\widetilde\theta_U-\theta^{\rm list}}\le\varepsilon/4$ with probability $1-e^{-As}$.

For each candidate support $V$ with $|V|\le s$, solve the finite linear feasibility system for $z$ supported on $V$ (with padded slots fixed to zero) satisfying
\[
 \mix{z-\widetilde\theta_U}\le\varepsilon/4
 \quad\text{for every }U\supseteq V,\ |U|\le2s.
\]
The true array is feasible on the high-probability event, and taking $U=V\cup S$ gives $\mix{z-\theta^{\rm list}}\le\varepsilon/2$ for any feasible $z$ (returning zero if infeasible). The design uses $2JpDq$ observations and at most $KpDq$ settings.
\end{proof}

Taking $p=m$ and selecting coordinatewise maximizers proves the predetermined-order upper bound in Corollary~\ref{thm:extremes}, since assignment regret is at most $2\mix{z-\theta}$.

\begin{lemma}[Known-block and individual validation]\label{lem:validation}
For at most $c s$ known blocks, all their labels can be estimated with mixed error at most any fixed fraction of $\varepsilon$, failure at most $e^{-As}$, and $O_{A,c}(sm(1+\chi H/b))=O_{A,c}(n_b\chi)$ queries. An arbitrary list of at most $C D_*$ individual cells can instead be validated in one batch with $O_{A,\zeta}(D_*\chi\log(eH))$ queries so that every selection of at most $2s$ validation errors has total absolute value below any fixed fraction of $\varepsilon$.
\end{lemma}
\begin{proof}
Partition the known blocks into at most $\lceil cs/b\rceil$ groups of size at most $b$, pad each group to $P=2^{\lceil\log_2 b\rceil}$ with baseline, and measure positive and negative parts of each Sylvester Hadamard row $q_W=\lceil C_{A,c}(1+\chi H/b)\rceil$ times per slot. Orthogonality gives proxy $2\sigma^2\|v\|_2^2/(Pq_W)$ within each group and slot, and hence $O_c(\sigma^2s/(bq_W))$ for any mixed-norm test across at most $cs$ blocks. A union over $\exp(O_c(sH))$ tests yields the desired tolerance at query cost $O_c(sm q_W)=O_{A,c}(n_b\chi)$ since $b\le s$, $H=O(s)$, and $\chi\ge1$.

For a history-selected list of at most $CD_*=O(sH^4)$ cells, take $q_V=\lceil C_{A,\zeta}\chi\log(eH)\rceil$ single-cell--baseline pairs per cell and apply Lemma~\ref{lem:noise} over the $\exp(O(s\log(eH)))$ signed subsets of size at most $2s$. Hard list caps keep both query counts deterministic on every path.
\end{proof}

\section{A confidence-dependent information bound}\label{sec:confidence-frontier}
In this section allow any $h=\log(1/\delta)\ge\log16$, retaining
the other parameters in \eqref{eq:benchmark}. Denote the physical
query matrix on the Gaussian zero instance by $A_0$. Its rows index
queried cells, and its random length is the actual stopping time.
For $v$ with $v_{i,0}=0$ define
\[
 w(v)=\sum_i\big(\max_a v_{i,a}-\min_a v_{i,a}\big).
\]

\begin{lemma}[Directional energy at the zero instance]\label{lem:energy}
Every uniformly PAC algorithm with independent Gaussian observations
satisfies, for every fixed nonzero $v$ supported on at most $s$ blocks,
\begin{equation}\label{eq:energy}
 \E_0\|A_0v\|_2^2\ge
 \frac{\sigma^2 w(v)^2}{2\varepsilon^2}
                \kl(1/2,\delta).
\end{equation}
For a fixed admissible action-independent noise sequence law independent
of the private tape, it also satisfies
$\Prob_0\{A_0v=0\}\le2\delta$.
\end{lemma}
\begin{proof}
Take $c>2\varepsilon/w(v)$ and consider the two effects $cv$ and
$-cv$. For every full assignment their regrets sum to $cw(v)$, so
their success sets are disjoint. Under the zero law at least one
alternative's failure set has probability at least $1/2$.
Under that alternative the same failure set has probability at most
$\delta$. Binary data processing lower bounds the stopped relative
entropy from the zero law by $\kl(1/2,\delta)$. For either sign the
Gaussian chain rule gives
\[
 \mathrm{KL}(P_0\Vert P_{\pm cv})
       =\frac{c^2}{2\sigma^2}\E_0\|A_0v\|_2^2 .
\]
This identity follows first for stopped finite prefixes and then by
monotone convergence of their information increments. If the energy
is infinite the bound already holds. Almost-sure termination makes
the output measurable on the stopped transcript. Let $c$ decrease to
$2\varepsilon/w(v)$. For the exact-zero assertion, couple both signs
to the zero run using the same noise and random tape. On $A_0v=0$
all observations, subsequent actions and outputs agree by induction.
At least one alternative fails there. The sum of the two marginal
failure probabilities is at most $2\delta$.
\end{proof}

Choose independently of the zero transcript a uniform set of $s$
blocks, one uniform label in each, and independent symmetric signs
on those cells. Then $w(v)=s$. Averaging the square of each row sum
over the signs kills its cross terms. A row of width $d$ contributes
$ds/N$ on average. Tonelli's theorem at the zero stopping time and
\eqref{eq:energy} imply
\[
 \E_0 C_T\ge \frac{N\chi}{2s}\kl(1/2,\delta).
\]
Since $\kl(1/2,\delta)=\tfrac12\log(1/[4\delta(1-\delta)])=\Omega(h)$
for $h\ge\log16$, this supplies the confidence term in
Theorem~\ref{thm:uniform}. The limited-batch lower bound at
$\delta=1/16$ applies to every more accurate algorithm.

\section{A cost bound uniform in the number of batches}\label{sec:uniform}
Define $L_0(u)=u$ and $L_{r+1}(u)=\max\{1,\log L_r(u)\}$ for $u\ge1$.
We prove Theorem~\ref{thm:uniform} for $h\le s$ under \eqref{eq:benchmark}
with constants depending only on $\nu,\zeta$, uniformly over $B$, $b$, and $m$.

\subsection{A uniform lower bound}
Let $V=(Km)^{-1}\sup_\theta\mathbb E_\theta C_T$ in the Gaussian subclass,
and put $x=1+V/\chi$. In the absent-group construction of
Section~\ref{sec:lower}, Lemma~\ref{lem:reference} gives
\[
 P_0(w_{j+1}\ge u)\le 2V/u,
\]
and for $W_j=\sum_{i\le j}w_i$ and likelihood ratio $\Lambda_j$,
\eqref{eq:lr-tail} with $A_0=64$ gives
\[
 P_+(W_j\le\chi t,\ \log\Lambda_j>A_0t+u)\le e^{-u}.
\]
Because $w_{j+1}$ is measurable before batch $j+1$ is read, setting
$p_j=2^{-j-10}$, $t_0=0$, and
\begin{equation}\label{eq:uniform-recursion}
 t_{j+1}=t_j+x+1+2x p_j^{-2}e^{A_0t_j},\qquad j\ge0,
\end{equation}
bounds both the transferred probability of $w_{j+1}>\chi(t_{j+1}-t_j)$ on
$\{W_j\le\chi t_j,\ \Lambda_j\le e^{A_0t_j}/p_j\}$ and the likelihood
exception by $p_j$. Induction yields
\[
 P_+(W_B>\chi t_B)\le2\sum_{j\ge0}p_j<1/16
\]
for every $B\ge1$. Lemma~\ref{lem:need-exposure} therefore forces
$t_B>c_0H$ for $c_0=c_0(\nu)>0$.

From \eqref{eq:uniform-recursion}, $t_1\le a_1x$ with $a_1=2^{21}+2$ and
$t_j\ge x+j$ for $j\ge1$. Since $2p_j^{-2}=2^{21}4^j$, we have
$t_{j+1}\le e^{80t_j}$, and $y_j=a_3t_j$ with $a_3=128$ satisfies
$y_{j+1}\le e^{y_j}$ for $j\ge1$. Hence
\[
 a_3c_0H<y_B\le\exp^{(B-1)}(a_1a_3x).
\]
Applying $L_{B-1}$ gives $L_{B-1}(a_3c_0H)\le a_1a_3x$. Because
$u\mapsto\max\{1,\log u\}$ is $1$-Lipschitz on $[1,\infty)$, rescaling its
initial argument by a positive constant shifts every iterate by at most a
fixed additive constant (and scales linearly at $B=1$), giving
\[
 L_{B-1}(H)\le a_4(1+V/\chi)+a_5.
\]
Proposition~\ref{prop:coverage} ($V\ge c_1\chi$) absorbs the additive
constants to give $C_B^*=\Omega(Km\chi L_{B-1}(H))$, and $C_T\le bT$ gives
$T_B^*=\Omega(n_b\chi L_{B-1}(H))$.

\subsection{The schedule and its deterministic sums}
For $B=1$, apply Lemma~\ref{lem:fixed-validator} with $p=m$.
For $B\ge2$ and large $s$, let $B_0\le B$ be the largest integer with
$B_0\ge2$ and $\log^{(B_0-1)}H\ge2$ (taking $B_0=B$ if none reaches $2$).
Run $B_0-1$ compression batches and one final validation batch, leaving
remaining batches empty. For $1\le j<B_0$, set
\begin{equation}\label{eq:uniform-schedule}
 \begin{aligned}
 x_j&=\log^{(B_0-j)}H,& D_j&=\lceil C_Dx_j\rceil,\\
 \ell_j&=\lceil m e^{-4x_j}\rceil,&q_j&=\lceil C_Q\chi4^{j-1}\rceil,\\
 r_j&=\lceil s/2^j\rceil,&\tau&=\varepsilon/(A_\tau s),
 \end{aligned}
\end{equation}
with constants depending only on $\nu,\zeta$. Since $x_{j+1}=e^{x_j}\ge2x_j$,
$x_1\ge2$, and $x_{B_0-1}=\log H$,
\begin{equation}\label{eq:uniform-sums}
 \begin{gathered}
 2^{B_0-1}\le\log H,\qquad B_0=O(\log\log H),\\
 \sum_{j<B_0}D_j^{-1}=O(1),\qquad\sum_{j<B_0}r_j\le s+B_0,\\
 \sum_{j=2}^{B_0-1}4^{j-1}x_j^{-3}\le x_1^{-3}\sum_{j\ge2}2^{-(j-1)}=O(1).
 \end{gathered}
\end{equation}

Batch one combines Lemma~\ref{lem:screen} ($t_0=\varepsilon/(sH)$, producing
$W$ and $D_*=C_*sH^4$) with a full-slot table ($p_1=m$); stage $j\ge2$ uses
$p_j=2\ell_{j-1}$ slots per ordinary block. Call a row high when its paired
mean exceeds $\tau$ in magnitude. At $j=1$, call a row usable if it avoids
$W$, and retain a label outside $W$ if fewer than $D_1/2$ of its $D_1$ rows
are usable or at least $D_1/8$ usable rows are high. At $j\ge2$, retain a
real label if at least $D_j/8$ of its rows are high. Any block retaining
more than $2\ell_j$ labels becomes exceptional: remove it from subsequent
compression and reserve all $m$ of its labels for final known-block
validation, while discarded labels in ordinary blocks pass to the final
residual table. Abort and return baseline if $|W|>2s$ or more than $2r_j$
blocks become exceptional at stage $j$, keeping total known blocks at most
$2s+2\sum_{j<B_0}r_j\le6s$ for large $s$.

\subsection{Compression with uniform constants}
If more than $2r_j$ blocks become exceptional at stage $j$, at least $r_j$
blocks each retain $\ell_j$ labels under one rule (masking or high rows),
giving a witness $F$ of $u_j=r_j\ell_j$ cells across at most $r_j\le s$
blocks. Because Lemma~\ref{lem:forest} requires only $|\operatorname{blocks}(F)|\le K/2$,
it applies even when $u_j>K/2$. Uniformly over $j<B_0$, $\ell_j\ge m/H^4$
and $2^j\le\log H$ imply
\[
 \begin{gathered}
 u_j\ge\frac{c sm}{H^4\log H}\gg D_*+sH,
 \\ |\operatorname{blocks}(F)|\le s,
 \qquad \log(p_j/\ell_j)=\Theta(x_j),
 \end{gathered}
\]
with witness log-cardinality at most
$r_j\log(eK/r_j)+u_j\log(ep_j/\ell_j)\le C u_jx_j$.

For stage-one masking, \eqref{eq:block-hits} bounds the hit probability by $2^{D_1u_1}(8s/J)^{D_1u_1/2}$, which for large $C_J$ and $C_D$ absorbs both
witness entropy and $|\mathcal W_\theta|=\exp(O(s))$. For high-row
retention, \eqref{eq:forest} gives $|\Gamma_j(F)|\ge(1-\alpha)D_ju_j$
($\alpha<1/64$) with exponent $\Omega(D_ju_j\log(Jp_j/(Cu_j)))$ covering
the same entropy. Significant cells touch at most $D_jD_*=o(D_ju_j)$ rows,
and Lemma~\ref{lem:forest} allows at most $D_js=o(D_ju_j)$ rows with weak
sum above $\tau/2$, forcing $\Omega(D_ju_j)$ distinct rows with noise above
$\tau/2$; since $q_j\tau^2/\sigma^2\ge C_Q/A_\tau^2$, the signed-sum bound
pays $\exp(O(D_ju_j))$ plus witness entropy, bounding stage failure by
$e^{-As}$ uniformly in $b,B_0,j,m$.

Writing $\mathcal S_j$ for the significant cells still measured at stage $j$
(unioning over $W\in\mathcal W_\theta$ at $j=1$ and conditioning on history
at $j\ge2$), \eqref{eq:small-collisions} gives $a_{D_j}(\mathcal S_j)\le C_{\rm coll}s/H$
with failure $e^{-As}$ since $p_j\ge m/H^4$ and $D_j=O(\log H)$. Collecting
discarded significant cells with at least $D_j/8$ collision incidences into
$X_j$ gives
\[
 \left|\bigcup_{j<B_0}X_j\right|
 \le\frac{16C_{\rm coll}s}{H}\sum_{j<B_0}D_j^{-1}
 \le C_ks/H.
\]
Let $X$ be the subset of $\bigcup_{j<B_0}X_j$ outside the finally known
blocks, with $|X|\le k:=\lceil C_ks/H\rceil$. The remaining discarded
significant cells form disjoint stage tails $\mathcal T_j$ satisfying
\begin{equation}\label{eq:tail-union}
 \bigcup_{j<B_0}\mathcal T_j\subseteq S_1,
 \qquad\left|\bigcup_{j<B_0}\mathcal T_j\right|\le D_*.
\end{equation}

\subsection{Summable decision error}
Every cell in $\mathcal T_j$ has at least $D_j/4$ usable low rows free of
other significant cells; those singleton rows are disjoint across cells in
$\mathcal T_j$. Summing $|\xi_{j,r}|$ over the full neighborhood
$\Gamma_j(F)$ avoids conditioning on which rows are low, and partitioning
$\Gamma_j(F)$ into groups of at most $D_js$ rows for the weak-load bound of
Lemma~\ref{lem:forest} yields, for every $F\subseteq\mathcal T_j$,
\begin{equation}\label{eq:uniform-discard}
 \tfrac14\|\theta_F\|_1
 \le\tau|F|+D_j^{-1}\sum_{r\in\Gamma_j(F)}|\xi_{j,r}|
       +\frac{C t_0H}{\log(eH)}(|F|+s).
\end{equation}

Let $O_j$ be the active-block positive optima discarded into $\mathcal T_j$
outside finally known blocks. Candidate sets and row signs have log size
$O(D_js)$, so Lemma~\ref{lem:noise} gives
\[
 D_j^{-1}\sum_{r\in\Gamma_j(O_j)}|\xi_{j,r}|
 \le C\varepsilon\sqrt{\chi/q_j}
 \le\frac{C\varepsilon}{\sqrt{C_Q}}2^{-(j-1)}.
\]
Since $\sum_j|O_j|\le s$, summing \eqref{eq:uniform-discard} gives
\begin{equation}\label{eq:uniform-best-loss}
 \sum_{j<B_0}\|\theta_{O_j}\|_1
 \le\frac{4\varepsilon}{A_\tau}
       +\frac{C\varepsilon}{\sqrt{C_Q}}
       +C\varepsilon B_0/\log(eH)
 \le\varepsilon/8
\end{equation}
for large $A_\tau,C_Q$ and $s\ge s_0(\nu,\zeta)$. Similarly, for any fixed
$C_{\rm loc}$ and $f_0=\lceil C_{\rm loc}s/H\rceil$, subsets of size at most
$f_0$ in $\mathcal S_j$ plus row signs and $\mathcal W_\theta$ have log size
$O(s)$, giving stage noise $O(\varepsilon H^{-1/2}2^{-(j-1)})$ and hence
\begin{equation}\label{eq:uniform-tail-mass}
 \sum_{j<B_0}\|\theta_{F_j}\|_1
 \le C\varepsilon\left(H^{-1/2}+H^{-1}+B_0/\log(eH)\right)
 =o(\varepsilon)
\end{equation}
uniformly over all $F_j\subseteq\mathcal T_j$ with $\sum_j|F_j|\le f_0$;
constants in \eqref{eq:uniform-tail-mass} may depend on $C_{\rm loc}$,
whereas the schedule \eqref{eq:uniform-schedule} does not. Since
$B_0=O(\log\log H)\le H$, unioning all stage failures stays below $e^{-2s}$.

\subsection{One final batch}
The final batch jointly executes known-block Hadamard validation on at most
$6s$ blocks (Lemma~\ref{lem:validation}), fixed-list validation on at most
$2\lceil m/H^4\rceil$ surviving labels per ordinary block
(Lemma~\ref{lem:fixed-validator}), and a fresh degree-$D_v$ residual table
certified through $2k$ cells ($\lceil Cs\rceil$ tries, $q_v=\lceil C_v\chi\rceil$
pairs per row) on discarded labels outside known blocks.

Writing $\mathcal T=\bigcup_j\mathcal T_j$, for any $|F|=f\le2k$,
\eqref{eq:factorial-hits} bounds light-block hits $Z$ from $\mathcal T\setminus F$
by $\mathbb E\binom Zq\le\mu_f^q/q!$ with $\mu_f=CD_v|\mathcal T|f/(Jm)\le CfH^4/m$,
while $F$ and heavy blocks add at most $3f$ tail cells deterministically. At
$v_f=\lceil M(f+s/H)\rceil$, $(e\mu_f/v_f)^{v_f}\le e^{-\nu Hv_f/2}$ pays for
the $e^{(\zeta+2)Hf}$ choices of $F$ with failure $e^{-As}$ (doubled after
conditioning on layout certification), so every $\Gamma_v(F)$ touches at most
$C_{\rm near}s/H$ tail cells.

Choosing $C_{\rm loc}\ge C_{\rm near}+C_k$, \eqref{eq:uniform-tail-mass},
weak-load control $O_\zeta(\varepsilon/\log(eH))$, and Lemma~\ref{lem:noise}
($O(\varepsilon/\sqrt H)$ over entropy $O(kH+D_vk)=O(s)$) make $\theta_X$
feasible for
\[
 \|(y-A_vz)_{\Gamma_v(F)}\|_1\le\gamma D_v\varepsilon
 \quad\text{for every }|F|\le2k,\qquad |\operatorname{supp}z|\le k,
\]
with $\gamma=(1-2\alpha)/32$. By \eqref{eq:rip1}, any feasible $z$ satisfies
$\|z-\theta_X\|_1\le\varepsilon/16$, and \eqref{eq:uniform-tail-mass} plus
$3kt_0$ bounds $\sum_{c\in\operatorname{supp}z\setminus X}|\theta_c|$, giving
\[
 \|(z-\theta)_{\operatorname{supp}z\cup X}\|_1
 \le\|z-\theta_X\|_1+
       \sum_{c\in\operatorname{supp}z\setminus X}|\theta_c|
 \le\varepsilon/8.
\]
Setting the other two validators to mixed tolerance $\varepsilon/16$ and
breaking zero ties in favor of baseline yields total regret at most
$(\varepsilon/8+3st_0)+2(\varepsilon/16+\varepsilon/16+\varepsilon/8)<\varepsilon$
with total failure probability at most $e^{-s}\le\delta$.

\subsection{Uniform cost and all execution paths}
Batch one uses $O(n_b\chi x_1)$ observations. For $j\ge2$,
$p_j=O(mx_j^{-4}+1)$ gives
\[
 T_j\le Cn_b\chi\,4^{j-1}x_j^{-3}
          +C(K/b)\chi\,4^{j-1}x_j,
\]
summing to $O(n_b\chi)+O((K/b)\chi4^{B_0}\log H)=O(n_b\chi)$ by
\eqref{eq:uniform-sums} and $m\ge s^\nu$. Final validation costs
$O(n_b\chi)+O((K/b)(m/H^4+1)\chi H)=O(n_b\chi)$. Because aborts occur only
at batch boundaries or before reading an uncertified table,
$T\le Cn_b\chi x_1$ and $C_T\le bT\le CKm\chi x_1$ hold on every path, with
$x_1=\Theta(L_{B-1}(H))$ both when $B_0=B$ and when saturation gives
$2\le x_1<e^2$. Section~\ref{sec:all-confidence} extends the proof to
$h>s$.

\section{The statistical frontier at all confidence levels}
\label{sec:all-confidence}
Retain the size, capacity, and noise assumptions in \eqref{eq:benchmark},
and allow any $h\ge\log16$. Set $t=\max(s,h)$, $R=t/s$, and
$U_B=\max\{L_{B-1}(H),h/s\}$. We prove
\begin{equation}\label{eq:all-confidence-pair}
  T_B^*=\Theta(n_b\chi U_B),\qquad
  C_B^*=\Theta(N\chi U_B),
\end{equation}
uniformly in $B$ and $h$. Confidence-independent lower bounds follow from
$\delta=1/16$ in Section~\ref{sec:uniform}, and the $h/s$ lower bound
follows from \eqref{eq:energy}. It remains to prove the matching upper bound.

\paragraph{The fixed design.}
Fixing the block matrix of Lemma~\ref{lem:fixed-validator} by search before
any readings with $D=\lceil C_DH\rceil$ and $q=\lceil C\chi(1+R/H)\rceil$
gives mixed-test exponent $\Omega(sqD/\chi)$, which pays the $O(sH)$ support
and sign entropy plus $At$. With $p$ slots this costs $O((K/b)p\chi(H+R))$
observations and at most $b$ times as many settings, covering both $B=1$ (at
$p=m$) and all $B\ge1$ whenever $R\ge c_0 H$ for any fixed $c_0>0$.

\paragraph{Intermediate confidence.}
Suppose $1\le R<c_0H$ with a small fixed $c_0$ chosen below. Retain
$x_j,D_j,\ell_j,r_j,\tau$ from \eqref{eq:uniform-schedule} and set
\begin{equation}\label{eq:conf-precision}
 q_j=\left\lceil C_Q\chi\,4^{j-1}(1+R/D_j)\right\rceil ,
\end{equation}
so $D_j q_j=O(\chi4^{j-1}(D_j+R))$ is additive in $R$.

In batch one keep the screen's $t_0=\varepsilon/(sH)$,
$\lfloor m/H^4\rfloor$ sampled labels, $\Theta(H)$ passes, and
$\lceil C\chi H^2\rceil$ pairs, analyzed at $k_0=\lceil t/H\rceil$ bad
sampled cells and remaining significant cap $D_*'=C_*tH^4$. Because a
$k_0$-cell witness has log entropy $O(k_0H)=O(t)$, $\Theta(Hk_0)$
incidences, and forest base $H^{O(1)}/m$ ($k_0=O(s)$), increasing pass and
precision constants pays exponent $At$ in both the red--blue collision and
noise branches, as well as the sampling tail when significant labels exceed
$D_*'$. For $c_0<1/2$ and large $s$, $|W|\le s+k_0\le2s$ and
$\log|\mathcal W_\theta|=O(t)$.

Since $t\le sH$, Lemma~\ref{lem:forest} pays exponent $At$ on any $CDs$ rows
(and on $Cs$ rows at constant degree). Exceptional-block witnesses still
contain at least $csm/(H^4\log H)\gg D_*'+t+sH$ cells ($D_*'=O(sH^5)$,
$m\ge s^\nu$), so the stage-one masking, forest, and high-row bounds pay
both $|\mathcal W_\theta|$ and $At$, while later layouts condition on
history. Applying the significant-set forest bound at collision excess
$q_c=\lceil C_{\rm coll}t/H\rceil$ (base $CD_j(D_*')^2/(q_cJp_j)\le H^{O(1)}/m$)
and summing $1/D_j$ bounds the exceptional discarded union by
$k=\lceil C_kt/H\rceil$.

\paragraph{Loss and local interference.}
On omitted positive optima, normalized stage noise remains at most
$C\varepsilon\sqrt{(\chi/q_j)(1+R/D_j)}\le(C/\sqrt{C_Q})\varepsilon2^{-(j-1)}$,
keeping their total loss at most $\varepsilon/8$. For fixed $C_{\rm loc}$ and
$f_0=\lceil C_{\rm loc}t/H\rceil$, subsets up to $f_0$ plus row signs,
$\mathcal W_\theta$, and confidence exponent $At$ cost $O(t)$ in log
probability ($D_j=O(\log H)$). Dividing Lemma~\ref{lem:noise} by $D_j$ and
using $R/(D_j+R)\le1$ bounds normalized neighborhood noise by
$C_{\rm loc}'\varepsilon 2^{-(j-1)}\sqrt{R/(HC_Q)}$, and summing thresholds,
stage noises, and weak loads yields
\begin{equation}\label{eq:bridge-mass}
 C_{\rm loc}''\varepsilon
 \left(\sqrt{R/H}+R/H+B_0/\log(eH)\right).
\end{equation}
Choosing $c_0$ sufficiently small after fixing the geometric and precision
constants makes \eqref{eq:bridge-mass} an arbitrarily small fixed fraction
of $\varepsilon$.

In the final batch, known-block Hadamard validation uses
$\lceil C(1+\chi(H+R)/b)\rceil$ pairs per signed row ($O(n_b\chi)$ for
$R\le H$), surviving-list validation costs $O((K/b)(m/H^4+1)\chi(H+R))$, and
the degree-$D_v$ residual table uses $\lceil Ct\rceil$ certification tries
through $2k=O(s)=o(K)$ and $q_v=\lceil C_v\chi R\rceil$ pairs per row. At
$v_f=\lceil M(f+t/H)\rceil$, light-block occupancy mean
$CD_vRH^4f/m$ gives tail base $e^{-\nu H/2}$, paying $N^f$ tests and $At$ so
every neighborhood touches $O(t/H)$ tail cells. Normalized residual noise is
$O(\varepsilon\sqrt{R/(HC_v)})$ over entropy $O(kH+D_vk)=O(t)$, yielding the
same $\varepsilon$-PAC comparator bound as in Section~\ref{sec:uniform} with
total failure at most $e^{-h}$.

\paragraph{Budget and round threshold.}
Batch one costs $O(n_b\chi(x_1+R))$, later compression tables sum via
$\sum 4^{j-1}x_j^{-3}=O(1)$ and $\sum 4^{j-1}x_j^{-4}=O(1)$ with rounding
$O((K/b)\chi(\log H)^2(\log H+R))=o(n_b\chi(1+R))$, and the residual table
costs $O(n_b\chi R)$. Since $x_1=\Theta(L_{B-1}(H))$, \eqref{eq:all-confidence-pair}
holds pathwise. For $z=\max\{2,h/s\}$ and $j_z(H)=\min\{j\ge0:L_j(H)\le z\}$,
$1+j_z(H)$ batches achieve the adaptive floor $\Theta(n_b\chi R)$ and
$\Theta(N\chi R)$; conversely, matching either floor within factor $L$
forces $L_{B-1}(H)\le C_Lz$ and hence $B\ge1+j_z(H)-O_{\nu,\zeta,L}(1)$.

\section{A computation bound that uses the available query budget}
\label{sec:budget-compute}
This section constructs the parallel multiscale filter for
Theorem~\ref{thm:adaptive-compute} and proves \eqref{eq:runtime};
Section~\ref{sec:res-cert} supplies the polynomial layout certificates.
Multiple low-precision tables are opened in parallel up to the degree budget
permitted by $\chi$, synchronized across shared batch boundaries.

\subsection{Repeated collisions of a prescribed set}
\begin{lemma}\label{lem:bad-incidences}
Let $G$ be a fixed set of $f\le D_*=C_*sH^4$ cells supported in
at most $s$ blocks of a full-slot balanced layout with $d$ passes.
Call a cell of $G$ bad if in at least $d/4$ passes it shares a row
with another cell of $G$. For fixed sufficiently large $d_0$ and all
$d\ge d_0$, there is $c_\nu>0$ such that
\begin{equation}\label{eq:bad-incidences}
 \Prob\{\#\text{ bad cells}\ge r\}\le e^{-c_\nu dHr},
 \qquad 1\le r\le f .
\end{equation}
Here $c_\nu$ can be chosen independently of subsequent increases in
$d_0$.
\end{lemma}
\begin{proof}
Fix $U\subset G$ of size $r$. In each pass, in every collision row meeting
$U$, connect its $U$-cells to a partner in $G\setminus U$ if one is present,
and otherwise join its $U$-cells in a spanning tree. A row with $t$
collision incidences from $U$ contributes at least $t/2$ forest edges with an
endpoint in $U$. When all cells of $U$ are bad, the passwise forests supply
at least $q=\lceil dr/8\rceil$ edges out of at most $\binom{drf}{q}$ edge
choices, spanning at most $s$ blocks. By the forest calculation of
Lemma~\ref{lem:forest}, a union over $U$ gives probability at most
\[
 \binom fr \left(\frac{C'f}{Jm}\right)^q.
\]
Since $J\ge cs$, $m\ge s^\nu$, and $f\le C_*sH^4$, $\log(C'f/(Jm))\le-\nu H/2$
for large $s$, while $\log\binom fr=O(Hr)$; choosing $d_0$ large once proves
\eqref{eq:bad-incidences}.
\end{proof}

\subsection{Schedule, decoding lists, and caps}
Run Lemma~\ref{lem:screen} in batch one; on success, condition on the screen
history with $G=S_1$ ($f=|G|\le D_*$) and mask all cells of $W$. Write
$g=1+h/H$ and set
\begin{align}
 t&=\left\lceil\log_2\min\{\chi,g\}\right\rceil,\qquad
 d_j=d_0\,2^j,\quad a_j=2^{3j/4},\quad 0\le j\le t,
 \label{eq:depths}\\
 \tau_j&=\frac{\varepsilon a_j}{A_\tau s},\qquad
 q_{j\ell}=\left\lceil C_Q\chi x_\ell/a_j^2\right\rceil ,
 \label{eq:band-precision}
\end{align}
with integer $d_0$ divisible by $8$ and $d_0\ge4A_0$. Draw independent
balanced layouts $j=0,\ldots,t$ ($n_j=d_jJm$ rows) before filtering. With
$A_0$ large relative to $c_\nu^{-1}$, set $k_j=\lfloor A_0g/d_j\rfloor$ and
$k=k_t$; by Lemma~\ref{lem:bad-incidences} at $r=k_j+1$, every layout has at
most $k_j$ bad cells of $G$ except with probability at most
$(t+1)e^{-c_\nu A_0(h+H)}\le\delta/16$ (and $h\le\chi H$ already gives $k=0$).

Define levels $x_0,x_1,\ldots,x_L=H$ from a large constant $x_0$ via
$x_{\ell+1}=\min\{H,\exp(x_\ell/2)\}$, so $L+1\le\log^*H+O(1)$,
$\sum_\ell x_\ell=O(H)$, and $\sum_\ell x_\ell^{-1/2}=O(1)$. At batch level
$\ell$, measure every active row of layout $j$ with $q_{j\ell}$ fresh
row--baseline pairs; keep a row active while its paired mean exceeds $\tau_j$
in magnitude, saving its exiting mean otherwise, and mark rows surviving
level $L$. Enforce row caps
\begin{align*}
 w_{j\ell}&=d_jD_*+2s+\lceil n_j e^{-x_\ell}\rceil\quad(\ell<L),\\
 w_{jL}&=d_jD_*+2s ,
\end{align*}
returning baseline at the batch boundary if any cap is exceeded. Place a cell
outside $W$ into list $j$ if at least $d_j/2$ of its rows are marked, abort
to baseline if any list reaches $F_*=\lceil16(D_*+2s+1)\rceil$, and let
$\mathcal U$ be the union of lists.

At most $d_jD_*$ rows meet $G$, and since $d_j=O(1+s/H)$ and $\log n_j=O(H)$,
Lemma~\ref{lem:forest} caps weak sums on any $s$ rows at
$O(t_0sH/\log(eH))=o(\varepsilon)$, leaving fewer than $s$ weak-only rows
above $\tau_j/2$. Exceeding $w_{j\ell}$ therefore requires
$r=s+\lceil n_je^{-x_\ell}\rceil$ ($\ell<L$) or $r=s$ ($\ell=L$) distinct
rows with noise above $\tau_j/2$; since $q_{j\ell}\tau_j^2/\sigma^2\ge C_Qx_\ell/A_\tau^2$
and $\log(en_j/r)$ is at most $1+x_\ell$ ($\ell<L$) or $O(H)$ ($\ell=L$),
signed-noise union bounds keep cap failures below $e^{-As}$. Likewise, any
$F_*$ listed cells touch at most $w_{jL}+d_jF_*/2$ rows and hence have
collision excess above $d_jF_*/4$; taking the union of
$(CF_*/(Jm))^{d_jF_*/4}$ over $(eN/F_*)^{F_*}$ candidate sets before
observing marks keeps list-cap failure below $e^{-As}$ without an algorithmic
expansion test.

\subsection{The mass of omitted optima and of the final tail}
Fix one positive optimum per active block having one, and let $E_j\subseteq G$
be those first having fewer than $d_j/4$ collisions in layout $j$, so
\begin{equation}\label{eq:assigned-size}
 |E_0|\le s,\qquad |E_j|\le k_{j-1},\qquad
 d_j|E_j|\le 2A_0g\quad (j\ge1).
\end{equation}
Any cell in the omitted subset $M_j=E_j\setminus\mathcal U$ has at least
$d_j/4$ unmarked rows free of other significant cells, disjoint across
$M_j$, each exiting with saved mean at most $\tau_j$:
\begin{align}\label{eq:assigned-loss}
 \|\theta_{M_j}\|_1
 &\le 4\tau_j|E_j|+
 \frac4{d_j}\sum_{r\in\Gamma_j(E_j)}
       \left(|b_{jr}|+\sum_{\ell=0}^L|\xi_{j\ell r}|\right),
\end{align}
where $b_{jr}$ is the weak-effect sum in row $r$. Applying
Lemma~\ref{lem:noise} at $v=d_j|E_j|=O(s)$ bounds normalized level noise by
$O(\varepsilon/\sqrt{C_Qx_\ell})$ at $j=0$ and
$O(\varepsilon(a_j/d_j)\sqrt{g/(sC_Qx_\ell)})$ at $j\ge1$ ($g=O(s/H)$), with
weak terms $O(\varepsilon/[d_j\log(eH)])$. Because
$\sum_{j\ge1}a_j/d_j=d_0^{-1}\sum_{j\ge1}2^{-j/4}<\infty$, choosing $A_\tau$
and $C_Q$ large bounds $\sum_j\|\theta_{M_j}\|_1\le\varepsilon/8$ (plus at
most $3st_0$ from optima below $3t_0$).

Partition unlisted significant cells outside $W$ into layout-$t$ bad cells
$X$ ($|X|\le k$) and remaining tail cells $\mathcal T$ ($|\mathcal T|\le D_*$).
When $k\ge1$ ($d_t\le A_0g$), for any fixed $C_{\rm loc}$ and
$u=\lceil C_{\rm loc}(k+g/d_t+1)\rceil$ ($d_tu=O(g)$), unioning the same
low-singleton bound over all subsets of the prespecified significant set
$G$ of size at most $u$ costs $O(uH)=O(s)$ in log cardinality. For any
such subset lying in $\mathcal T$, the threshold, weak-load, and normalized
noise terms are respectively $O(\varepsilon a_tg/(s d_t))$,
$O(\varepsilon/[d_t\log(eH)])$, and
$O(\varepsilon(a_t/d_t)\sqrt{g/s})$ after summing levels. Since
$g/s=O(1/H)$ and $a_t/d_t=O(1)$, this gives
\begin{equation}\label{eq:last-tail}
 \sup_{V\subset\mathcal T,\ |V|\le u}\|\theta_V\|_1=o(\varepsilon).
\end{equation}

\subsection{The final validation batch}
Validate known blocks and $\mathcal U$ ($|\mathcal U|=O(sH^5)$) via
Lemma~\ref{lem:validation} at tolerance $\varepsilon/16$ each. When $k\ge1$,
draw a fresh degree-$d_t$ layout, mask known blocks and $\mathcal U$, and
measure $q_v=\lceil C_v\chi/d_t\rceil$ pairs per row. Enumerating expansion
through $2k$ succeeds with probability at least $1/2$ per try across at most
$\lceil(h+\log64)/\log2\rceil$ tries (or use the polynomial certificates of
Section~\ref{sec:res-cert} in their regimes). Truncating the convex fit of
Section~\ref{sec:res-cert} to its $k$ largest absolute entries recovers
$\theta_X$ within $\ell_1$ error $\varepsilon/16$, and \eqref{eq:last-tail}
plus $3kt_0$ gives error at most $\varepsilon/8$ on $X\cup\supp z$, yielding
total regret below $\varepsilon$ with failure probability at most $e^{-h}$.

\subsection{Pathwise cost and time}
Since $\sum_{j=0}^t d_j=O(\chi)$, $\sum_{j=0}^t d_j/a_j^2=O(d_0)$, and
$\sum_\ell e^{-x_\ell/2}=O(1)$, initial passes and geometric cap terms cost
$O(n_b\chi)$, while
\[
 \sum_{j=0}^t (d_j D_*+2s+1)
       \{L+1+C_Q\chi H/a_j^2\}
       =O\!\left(\chi H(D_*+s)\right)=o(n_b\chi),
\]
singleton validation costs $O(sH^5\chi\log(eH))=o(n_b\chi)$, and the
residual table costs $O(Jm(d_t+\chi))=O(n_b\chi)$, giving pathwise optimal
$T$ and $C_T\le bT$. Only support enumeration through $2k$ takes
superpolynomial work $\operatorname{poly}(s,m,T_{\max})\exp(C_\zeta kH)$;
since $k=0$ for $\chi\ge g$ and $kH\le A_0(h+H)/(d_0\chi)$ otherwise,
choosing $d_0$ large for any given $\eta>0$ proves \eqref{eq:runtime}
(with bounded rejection-sampling retries in a fair-bit model charging at most
$\delta/32$).

\begin{corollary}[A polynomial loss bound for every $\chi$]\label{cor:head}
For the same data and caps, omit the residual table and fit. If
$v_{(1)}\ge\cdots\ge v_{(K)}$ are the block optima in decreasing
order, this polynomial-time rule has, with probability at least
$1-\delta$,
\[
 \mathcal R_\theta(\widehat x)\le
 \varepsilon/2+\sum_{i=1}^{\min(k,K)}v_{(i)} .
\]
\end{corollary}
\begin{proof}
Omitted significant optima in $\bigcup_j M_j$ cost at most $\varepsilon/8$,
any remaining omitted positive optima lie among the at most $k$ bad cells of
layout $t$, and setting the two direct validators to tolerance
$\varepsilon/32$ bounds all remaining errors plus $3st_0$ below
$\varepsilon/2$ for large $s$.
\end{proof}

Section~\ref{sec:res-cert} further shows polynomial-time optimality under
one-sided negative bounds $\theta_{i,a}\ge-M\varepsilon/s$, including at
$h=s$ and $\chi=1$.

\paragraph{A failure of the unchecked residual-free filter.}
At $m=b=s$, take identically zero noise under the supplied valid positive
proxy $\sigma^2$ and place two cells with effects $+2\varepsilon$ and
$-2\varepsilon$ in distinct blocks. Neither is sampled by the screen with
probability $(1-\lfloor m/H^4\rfloor/m)^2$, leaving $W=\emptyset$. In each
full-slot pass they share a row with probability
$p_s=m^{-1}\sum_g n_g(n_g-1)/(K(K-1))=\Theta(1/(sm))$. If they collide in
every pass across all independent filtering tables, all incident row means
vanish, so omitting both the residual table and the pair test returns
baseline and incurs regret $2\varepsilon$ with probability at least
$(1-\lfloor m/H^4\rfloor/m)^2p_s^{\sum_j d_j}\ge\exp(-C\chi H)$. Hence that
unchecked shortcut fails whenever $h/(\chi H)\to\infty$, though this
counterexample applies only to that shortcut.

\section{Polynomial certificates and a convex residual decoder}
\label{sec:res-cert}
Under \eqref{eq:benchmark} ($h\le s$, $g=1+h/H$), the multiscale filter of
Section~\ref{sec:budget-compute} may stop at any terminal degree $d=d_0 2^t$
satisfying $\sum_{j\le t}d_j=O(\chi)$ and $d=O(g+1)$, leaving
$k=\lfloor A_0g/d\rfloor$ exceptional cells (when $k=0$, the two direct
validators already suffice). We construct polynomial-time layout certificates
and a convex residual decoder for $k\ge1$.

\begin{theorem}\label{thm:poly-cert}
With arbitrary signed, unbounded effects the two adaptive minimax
costs and the $\log^*H+O(1)$ batch cap are attained in polynomial
real-arithmetic time in either of the following regimes:
\begin{align}
 &g=O(\chi^2),\label{eq:quadratic-regime}\\
 &g=O(\chi^3),\qquad g^{2/3}b^2/N\le s^{-\gamma}
       \quad\text{for fixed }\gamma>0 .\label{eq:cubic-regime}
\end{align}
Constants may depend on $\nu,\zeta,\gamma$ and the fixed implied
multipliers. In particular for $m=b=s$ at $h=s$ it suffices that
$\chi=\Omega((s/H)^{1/3})$.
\end{theorem}

\subsection{Two efficiently checkable statistics}
For two distinct columns $a\ne a'$ in a balanced residual layout, let
$D(a,a')$ count their common rows and let $W(a,a')$ count distinct third
columns $c$ meeting $a$ in one pass and $a'$ in a different pass. Both
$u=\max_{a\ne a'}D(a,a')$ and $v=\max_{a\ne a'}W(a,a')$ are computable in
$O(N^3d^2)$ elementary operations before the residual batch is read.

Any fixed column pair collides in one pass with probability at most
$C/(Jm)$ (zero within the same block), independently across passes, so for
$l\ge1$,
\begin{equation}\label{eq:codegree-tail}
 \Prob\{u\ge l\}\le N^2\left(\frac{C d}{Jm}\right)^l.
\end{equation}
Since $d=O(1+s/H)$, the base is at most $s^{-\nu/2}$ for large $s$, so a
fixed $u_0=u_0(\nu,\zeta)$ yields $\Prob\{u\le u_0\}\ge7/8$.

For $W(a,a')$, expose the layers and group positions of $a,a'$ across all
passes, fix $l$ distinct third cells, and assign each an ordered pair of
distinct passes ($2l$ required incidences). A block with $p'\ge m-2$ free
layers in a pass satisfies $w$ compatible layer constraints with probability
$1/(p')_w\le(e/p')^w\le(C/m)^w$, and conditional on layers, $w'$ additional
endpoints in a layer with $n'\ge K-2$ free positions land in prescribed
groups with probability at most $(\lceil K/J\rceil)^{w'}/(n')_{w'}\le(C/J)^{w'}$.
Multiplying across blocks, layers, and passes gives $(C/(Jm))^{2l}$, and
summing over third-cell sets, pass pairs, and endpoints $(a,a')$ yields
\begin{equation}\label{eq:witness-tail}
 \Prob\{v\ge l\}\le
  \frac{N^2}{l!}\left(\frac{C N d^2}{(Jm)^2}\right)^l .
\end{equation}
Under \eqref{eq:cubic-regime} with $d=O(g^{1/3})$, the base is
$O(g^{2/3}b^2/N)\le s^{-\gamma/2}$ for large $s$, so a fixed
$v_0=v_0(\gamma,\zeta)$ gives $\Prob\{v\le v_0\}\ge7/8$ (with $u_0,v_0$
fixed before choosing $d_0$). Joint acceptance thus exceeds $1/2$ per draw;
across at most $\lceil(h+\log 64)/\log2\rceil$ tries, exhaustion has
probability at most $\delta/64$ and returns baseline without taking
observations, while conditioning on acceptance inflates unconditioned
geometric failure probabilities by at most two.

\paragraph{Pair certificate and a median.}
Under \eqref{eq:quadratic-regime}, set $t=\lceil\log_2\sqrt g\rceil$ so
$d=\Theta(d_0\sqrt g)=O(\chi)$, certify $u\le u_0$ with $d\ge16u_0 k$, and
let $m_c$ be the median of the $d$ residual row means incident to discarded
cell $c$. For any $|F|\le2k$ and $c\in F$, deleting the at most $3u_0k$ rows
meeting $(X\cup F)\setminus\{c\}$ leaves at least $d/4$ private rows whose
deviations from $\theta_{X,c}$ equal $(y-A\theta_X)_r$ and have magnitude at
least $|m_c-\theta_{X,c}|$, giving
\[
 \sum_{c\in F}|m_c-\theta_{X,c}|\le
 (4/d)\|(y-A\theta_X)_{\Gamma(F)}\|_1 .
\]
Retaining the $k$ largest $|m_c|$ and setting $F$ to the union of $X$ and
the selected cells ($|\Gamma(F)|\le2dk$) bounds $\ell_1$ error against
$\theta_X$ by $16\gamma_0\varepsilon$ under \eqref{eq:top-r-feasible}.

\subsection{Expansion from distinct third columns}
Within each row touched by a set $F$ of $f$ cells (multiplicity $w$), match
$\lfloor w/2\rfloor\ge(w-1)/2$ disjoint pairs and deduplicate across rows to
form a simple graph on $F$ with at least $a_d(F)/(2u_0)$ edges when
$u\le u_0$ ($u_0\ge1$). Because each vertex belongs to at most one matched
edge per row, any common neighbor of two vertices meets them in distinct
passes, bounding common neighbors by $v_0$. Counting length-two paths gives
\[
 \sum_x\binom{\deg(x)}{2}\le v_0\binom f2,\qquad
 \bar d(\bar d-1)\le v_0(f-1),
\]
for average degree $\bar d$, and hence
\begin{equation}\label{eq:quadratic-expansion}
 a_d(F)\le u_0 f(1+\sqrt{v_0 f}).
\end{equation}
At $\alpha=1/16$ and large $d_0$, \eqref{eq:quadratic-expansion} enforces
$a_d(F)\le\alpha d f$ for all $f\le c(\alpha,u_0,v_0)d^2$. Under
\eqref{eq:cubic-regime}, taking $t=\lceil\log_2(g^{1/3})\rceil$ gives
$d=\Theta(d_0g^{1/3})=O(\chi)$ and $2k\le2A_0g/d\le cd^2$ for large $d_0$,
certifying expansion through $2k$ in polynomial time.

\subsection{Controlling the largest residual rows}
For $k\ge1$, set $r=dk\in[A_0g/2,A_0g]$ and write
$\|e\|_{[r]}=\sum_{i=1}^{r}|e|_{(i)}$ for the sum of the $r$ largest
absolute entries. Conditional on a successful filtering history, for any
fixed $\gamma_0>0$ and large $s$,
\begin{equation}\label{eq:top-r-feasible}
 \|y-A\theta_X\|_{[r]}\le\gamma_0 d\varepsilon
\end{equation}
with the prescribed confidence slack. Indeed, before layout acceptance, fix
any $r$ rows ($r_i$ in pass $i$, touching $p_\lambda$ groups in layer
$\lambda$ with $\sum_\lambda p_\lambda=r_i$). Averaging group-hit weights
$2p_\lambda/J$ over distinct layer assignments via the symmetric-mean
inequality bounds the probability that $a$ prescribed tail cells hit those
rows in pass $i$ by $(2r_i/(Jm))^a$. Summing over $\mathcal T$ and passes,
the total incidence count $Z$ satisfies
\[
 \E\binom Zl\le\frac1{l!}
       \left(\frac{2|\mathcal T|\sum_i r_i}{Jm}\right)^l ,
\]
with mean $O(rH^4/m)$. At $l=\lceil Mr\rceil$, the tail $\exp(-\nu HMr/2)$
pays the $\binom{dJm}{r}=\exp(O(rH))$ row-set choices and target failure
exponent since $rH\ge c(h+H)$. Dividing by $d$ gives each touched tail cell
weight at most $1$ and total weight $O(r/d)=O(k)$, which
\eqref{eq:last-tail} bounds by $o(\varepsilon)$. Weak loads remain
$o(\varepsilon)$, conditioning on acceptance doubles failure bounds at most,
and Lemma~\ref{lem:noise} at $q_v=\lceil C_v\chi/d\rceil$ over entropy
$O(rH)$ bounds normalized noise by
$C(\sigma/d)\sqrt{r(h+H+rH)/q_v}=O(\varepsilon\sqrt{k/s})=o(\varepsilon)$.

\subsection{Polynomial decoding after certification}
On the discarded columns outside known blocks, solve
\[
  \min_z\|z\|_1\quad\text{subject to}\quad
          \|y-Az\|_{[r]}\le \gamma_0 d\varepsilon ,
\]
returning $z=0$ if infeasible. Under \eqref{eq:top-r-feasible}, $\theta_X$
is feasible, so $w=z-\theta_X$ satisfies $\|w_{X^c}\|_1\le\|w_X\|_1$.
Partitioning $w$ in decreasing magnitude into $k$-coordinate blocks
$S_0,S_1,\ldots$, expansion through $2k$ gives
$\|A_{S_0}w_{S_0}\|_1\ge(1-2\alpha)d\|w_{S_0}\|_1$ and bounds cross-block
hits into $\Gamma(S_0)$ by $a_d(S_0\cup S_i)\le2\alpha dk$, summing to at
most $2\alpha d\|w\|_1$ \cite{berinde2008}. Since $|\Gamma(S_0)|\le r$,
\[
 2\gamma_0d\varepsilon\ge\|Aw\|_{[r]}
 \ge (1-6\alpha)d\|w_{S_0}\|_1,\qquad
 \|w\|_1\le2\|w_{S_0}\|_1 .
\]
Truncating $z$ to its $k$ largest absolute entries at most doubles
$\|z-\theta_X\|_1\le\varepsilon/16$ for small $\gamma_0$, and
\eqref{eq:last-tail} plus $3kt_0$ completes the $\varepsilon$-PAC guarantee.

Writing $\|e\|_{[r]}=\min_{\lambda\ge0}\{r\lambda+\sum_i(|e_i|-\lambda)_+\}$
expresses the program as a polynomial-size linear program
($u_i+\lambda\ge\pm(y-Az)_i$, $r\lambda+\sum_i u_i\le\gamma_0d\varepsilon$,
$\lambda,u_i\ge0$) whose standard-form constraint matrix has integer entries
bounded by $\max(1,r)$. Interior-point methods depending only on the
constraint matrix \cite{vavasis1996,dadush2024} therefore solve it in
polynomial real-arithmetic operations regardless of right-hand-side
magnitudes. In detail, the arithmetic bound of Vavasis and Ye depends on
the dimensions and the logarithm of a matrix condition measure.
For this integral matrix the latter is polynomial in its encoding length
\cite[Proposition~2.1]{dadush2024}. Basis inverses are governed by
determinant ratios with nonzero integer denominators. This gives the stated
cost for feasibility and optimization and completes Theorem~\ref{thm:poly-cert}.

\paragraph{Unbounded negative outliers at unit noise scale.}
Suppose at most $C_-\chi$ negative cells have arbitrary magnitude and all
other cells satisfy $\theta_{i,a}\ge-M\varepsilon/s$ for fixed $C_-,M$, with
unrestricted positive effects. Certifying $u\le u_0$ on the terminal
filtering layout of \eqref{eq:depths} with $d_0\ge8u_0(C_-+1)$ and omitting
the residual LP already attains both adaptive optima in polynomial time
throughout \eqref{eq:benchmark} (including $\chi=1,h=s$). Indeed, for
$k\ge1$, $d\ge d_0\chi$ ensures that removing rows meeting exceptional
negatives deletes at most $u_0C_-\chi\le d/8$ rows per missed positive
optimum. Each missed optimum has at least $d/4$ unmarked outlier-free
rows, in a union of at most $dk$ rows. Summing the saved responses once
per row and dividing by $d$ bounds the threshold contribution by
$O(\varepsilon a_tk/s)$ and the saved noise by
$O(\varepsilon(a_t/d)\sqrt{g/s})$. Apply the largest-row occupancy bound
to the whole significant set $G$, fixed given the screen history. Its
$O(dk)$ incidences in these rows contribute at most
$O(M\varepsilon k/s)=o(\varepsilon)$ from moderate negatives after division
by $d$. The weak load is $O(\varepsilon/[d\log(eH)])$. Together with
the two direct validators this bounds the regret by $\varepsilon$.

\paragraph{A polynomial budget outside the matching region.}
Scheduling with an inflated valid proxy $\widetilde\chi=\max\{\chi,g^{1/3}\}$
at $m=b=s$ yields pathwise query cost $O(n_b\max\{\chi,g^{1/3}\})$ in
polynomial time across \eqref{eq:benchmark}. At $h=s$ and $\chi=1$, this
exceeds the statistical optimum by $O((s/H)^{1/3})$; attaining a constant
factor in polynomial time for arbitrary signed effects there remains open.

\section{Packing individual estimates into legal measurements}
\label{sec:orthogonal}
The following physical measurement primitive applies to a list selected
on the completed history, including a different number of labels in
each block. Section~\ref{sec:rank-packing} uses it for arbitrary dimensions.

\begin{lemma}[Packing and conditional precision]\label{lem:packing}
Let a list contain $l_i$ distinct real labels from block $i$, let
$l=\sum_i l_i>0$, and set
\[
 G=\max\{\max_i l_i,\lceil l/b\rceil\},\qquad
 P=2^{\lceil\log_2 b\rceil}.
\]
In one batch of $2PqG$ observations and $Pql$ settings one can
estimate all $l$ effects. Given the history and plan, the error vector
$e$ satisfies for every fixed real $v$ and $\lambda$
\begin{equation}\label{eq:orthogonal-mgf}
 \E[\exp(\lambda\langle v,e\rangle)\mid{\cal F}_{\rm plan}]
 \le \exp\!\left(\lambda^2\sigma^2\|v\|_2^2/(Pq)\right).
\end{equation}
The group count $G$ is minimal for a partition with at most $b$
cells and at most one label of each block per group.
\end{lemma}
\begin{proof}
Concatenate the lists by block and assign consecutive cells cyclically
to groups $1,\ldots,G$. Each block uses distinct groups because
$l_i\le G$. Their sizes differ by at most one and do not exceed $b$.
The two terms in the maximum are necessary lower bounds. Pad each
group to $P$ columns with zeros. For each row of the Sylvester matrix
$U_P$, measure its positive and negative parts as two assignments
and subtract their responses. Each real cell occurs in exactly one
of them. The intercept cancels. Take $q$ fresh pairs per row and
multiply the averaged response vector by $U_P^\top/P$. The identity
$U_P^\top U_P=P I$ gives the true group means. For a linear test $v$
on its real columns, the squared sum of all physical coefficients
is $2\|v\|_2^2/(Pq)$. Summing over groups and applying sequential
conditioning in \eqref{eq:mgf} proves \eqref{eq:orthogonal-mgf}.
Every cost follows by counting the physical assignments. An empty
list takes zero observations.
\end{proof}

\section{Rank elimination for arbitrary dimensions}\label{sec:rank-packing}
Here $1\le s,b\le K$, $m\ge2$, $\chi\ge1$ and $h\ge\log16$ are
arbitrary. With $H_m=\log(em)$ and $U^m_B=\max\{L_{B-1}(H_m),h/s\}$,
we prove Theorem~\ref{thm:rank-general}.

\begin{proof}[Lower bounds]
Use $s$ disjoint groups of $\lfloor K/s\rfloor$ blocks, ignoring any
remaining blocks. Their total $K'=s\lfloor K/s\rfloor$ satisfies
$K/2\le K'\le K$. In the absent-group argument a reference group
contains $\lfloor K/s\rfloor m$ candidate cells and its output selects
at most $\lfloor K/s\rfloor$ of them. Its null selection probability
is therefore at most $1/m$. Summing empty-group budgets before averaging
over cells bounds its reference exposure by $2\sup\E C_T/(K'm)$,
at most $4\sup\E C_T/(Km)$. Active selection still has probability
at least $5/8$. For every $m\ge2$, Pinsker gives the coverage rate
since $5/8-1/m\ge1/8$. For large $m$, the stopped likelihood test
at $\sqrt m$ forces exposure of order $\chi\log m$ on an event of
constant probability. The summable transfer and the scaled exponential
recursion in Section~\ref{sec:uniform} now use $H_m$ in place of
$H$ and a numerical change in the reference-budget constant.
The additive constants after iterating the logarithm are absorbed
by coverage. Bounded $m$ follows directly by coverage. The reference
noise tape argument for a predetermined order still gives its full
$\log m$ factor even with stopping. Finally, averaging the energy
bound \eqref{eq:energy} over signed sparse directions applies to all
$K\ge s$ and supplies the term $h/s$.
\end{proof}

\begin{proof}[Algorithm and upper bound]
Initially keep all $p_0=m+1$ labels per block, including the noiseless
baseline. Put $z=\max(1,h/s)$ and
$x_1=\max\{2,z,L_{B-1}(H_m)\}$. If $x_1\ge H_m$ use one terminal
stage at $x_1$. Otherwise use $x_{j+1}=\min\{H_m,e^{x_j}\}$,
ending at $H_m$. Write $J_0$ for the last stage. Monotonicity of the
clipped logarithm implies $J_0\le B$. Scales at least double except
possibly at the last step.

In stage $j$ use the cyclic groups of Lemma~\ref{lem:packing} on all
remaining real cells. With $P=2^{\lceil\log_2 b\rceil}$ use
$q_j=\lceil C_Q\chi4^{j-1}x_j/P\rceil$ pairs per signed row.
Assign estimate zero to any surviving baseline. In each block keep
the $p_j$ largest estimated labels with a fixed tie rule, where
\[
 p_j=\lceil(m+1)e^{-4x_j}\rceil\quad(j<J_0),\qquad p_{J_0}=1 .
\]
The sizes decrease weakly. A baseline label may be discarded like
any other label. If no real cells remain, the decisions are already
fixed. All batch lengths are determined before their observations.

In a fixed active block, conditional on the preceding history, let
$a_i$ be a true maximizer in its current list. Write $e_{ia}$ for
the estimation errors and $\Delta_i$ for the decrease of the true
best value after retaining $p_j$ labels. If $a_i$ is discarded, every
retained estimate is at least its estimate. Averaging over these
labels and bounding their true values by the best retained value gives
\[
 0\le\Delta_i\le |e_{ia_i}|+
 \frac1{p_j}\sum_{\text{largest }p_j\text{ in block }i}|e_{ia}|.
\]
The inequality also holds when $a_i$ survives. A fresh current
maximizer is used at each stage, so these decreases telescope
exactly to the final regret.

Use \eqref{eq:orthogonal-mgf} for signed tests on the fixed true
active blocks. Tests for the first term have squared norm at most
$s$ and sign entropy at most $s\log2$. For the second term the squared
norm is at most $s/p_j$ and log cardinality at most
$s p_j\log(2e p_{j-1}/p_j)$. A baseline contributes no noise.
With conditional failure at most $2e^{-u_j}$, $u_j=h+j+3$, the sum
of the true best-value drops is at most
\[
 C\varepsilon\sqrt{\frac{\chi}{Pq_j}}
 \left(\sqrt{1+u_j/s}+
 \sqrt{\log(2ep_{j-1}/p_j)+u_j/(s p_j)}\right).
\]
Here $\log(2ep_{j-1}/p_j)=O(x_j)$, including at the terminal
stage. Also $u_j/s=O(x_j)$ since $x_1\ge z$ and $j=O(x_j)$.
Thus the $j$th error is at most
$C\varepsilon 2^{-(j-1)}/\sqrt{C_Q}$. Choose $C_Q$ to pay their
sum and note that $2\sum_{j\ge1}e^{-u_j}\le e^{-h}$.

The number of groups at stage $j$ is at most
$2Kp_{j-1}/b$. Thus the initial cost is $O((Km/b)(b+\chi U^m_B))$.
For $j\ge2$, $p_{j-1}\le(m+1)x_j^{-4}+1$, also when the final step
is truncated. The cost proportional to $(m+1)x_j^{-4}$ sums using
$\sum_{j\ge2}4^{j-1}x_j^{-3}=O(1)$. The one-slot rounding part
costs $O((K/b)(bJ_0+\chi H_m^3))$ because $2^{J_0-2}\le H_m$
and $\sum_j4^{j-1}x_j=O(H_m^3)$ for the multistage case.
Both $J_0$ and $H_m^3$ are $O(m)$ uniformly for $m\ge2$.
These estimates give the observation bound using all of the
available capacity. To attain the stated setting cap simultaneously,
run the same algorithm with working width
$w=\min\{b,\lceil\chi U_B^m\rceil\}$. Every action remains legal.
We now have $T=O((Km/w)(w+\chi U_B^m))$, which obeys the displayed
observation bound in both cases for $w$, and
$C_T\le wT=O(Km\chi U_B^m)$ universally. Packing, transforms and
sorting take polynomially many arithmetic and index operations.
\end{proof}

\paragraph{A scalar information limit at wide capacity.}
For general dimensions, every adaptive Gaussian algorithm additionally
satisfies
\[
 \sup_\theta\E_\theta T\ \ge\
 \frac{c s\log m}{\log(1+16\min(b,s)^2/\chi)} .
\]
Use a fixed support of $s$ blocks with independent uniform best
labels of value $8\varepsilon/s$. Truncate at the integer part of a
large fixed multiple of the worst expected reading count, and pad
after early stopping. The Fano argument above requires
$\Omega(s\log m)$ nats. Given any preceding history and seed, the
next signal mean ranges over an interval of length at most
$8\varepsilon\min(b,s)/s$. Its variance is at most a quarter of
the squared range. Gaussian maximal entropy bounds the conditional
mutual information in one read by half the denominator. Summing over
the virtual prefix proves the claim, including feedback stopping.
For predetermined orders use a uniform support $U$ given only to the
analysis decoder. In a virtual row meeting $d_U$ support blocks, the
label-averaged signal variance is at most
$(8\varepsilon/s)^2 d_U/m$. Entropy subadditivity followed by Jensen
over $U$, with $\E d_U\le sb/K$, replaces the denominator above by
$\log(1+64sb/(Km\chi))$. In particular $K=s=b$, $m=2$, $\chi=1$
and fixed confidence require $\Omega(s/\log(es))$ scalar readings
even though the optimal setting count divided by capacity is $O(1)$.

\paragraph{Integer schedules in the arithmetic model.}
The polynomial-time statements can use dyadic schedules. Obtain an
upper dyadic approximation to $h$ within a constant factor by
successively squaring $1/2$ and comparing with $\delta$. Replace
the terminal log scale by $1+\lceil\log_2 m\rceil$ (or
$1+\lceil\log_2 s\rceil$ for the row filter). In the rank schedule
use the backward map $D(u)=\max(2,\lceil\log_2u\rceil)$ and the
forward map $2^x$, truncated at the known terminal scale. The forward
map inverts each backward step with upward slack. If $u\le A v$,
then $D(u)\le A\max(1,\log v)$ for a sufficiently large fixed $A$
and $v\ge1$, by $\lceil\log_2(Av)\rceil\le1+\log_2 A+\log v/\log2$.
Induction preserves a constant-factor comparison with every clipped
iterate. Use $2^{-4x}$ for list shrinkage. In the shared row filter
use $\min(H_2,2^{\lfloor x/2\rfloor})$ and $2^{-x}$ in the caps,
with terminal integer $H_2$. The product for the next-stage cost
remains at most $2^{-x/2}$. After a fixed rescaling this recursion
grows at least as fast as $y\mapsto e^y$ until truncation. Ceilings
of repetition counts can be replaced by upward powers of two, with
at most twice the cost. Threshold comparisons, integer searches and
repeated squaring implement these choices in polynomial work.

\section{Finite decoder and measurement identities}\label{sec:finite-codes}

\subsection{When the residual fit vanishes}
Write $J_r(e)$ for the sum of the $r$ largest absolute coordinates.
Let $L_i$ be the directly validated labels in block $i$, including any
previously discovered block, and always allow label zero.
Define the restricted-optimum loss
\[
 D_L(\theta)=\sum_{i=1}^K
 \bigl(\max_{0\le a\le m}\theta_{ia}
       -\max_{a\in L_i\cup\{0\}}\theta_{ia}\bigr).
\]
\begin{proposition}\label{prop:vanishing-fit}
For a nonnegative radius $\eta$, the zero vector is a minimizer of
$\min\|z\|_1$ subject to $J_r(y-Az)\le\eta$ if and only if
$J_r(y)\le\eta$, in which case $z=0$ is the unique minimizer. If the
decision defaults to baseline on ties at zero and otherwise uses the
direct estimates and this residual fit, its loss is at least
$D_L(\theta)$ whenever $J_r(y)\le\eta$.
\end{proposition}
\begin{proof}
Substitution gives feasibility of $z=0$, which uniquely attains
$\|z\|_1=0$. When $z=0$, every chosen label belongs to $L_i\cup\{0\}$,
giving $\mathcal R_\theta(\widehat x)\ge D_L(\theta)$ block by block.
\end{proof}
Testing $J_r(y)\le\eta$ uses only observations, whereas $D_L(\theta)$ is an
oracle diagnostic; a nonzero residual fit can deploy labels outside $L$ and
achieve loss below $D_L(\theta)$.

\paragraph{What a residual budget can recover.}
Let $d_i$ be block $i$'s summand in $D_L(\theta)$ and order these gaps as
$d_{(1)}\ge\cdots\ge d_{(K)}$. Adding at most $k$ cells closes at most $k$
block gaps, giving exact oracle floor
\[
 D_{L,k}=\min_{|B|\le k}D_{L\cup B}
         =\sum_{j>\min(k,K)}d_{(j)}.
\]
For any output whose chosen nonbaseline labels outside $L$ form $B$ with
$|B|\le k$ (such as a $k$-truncated residual fit), realized loss decomposes
into three nonnegative terms:
\[
 \mathcal R=D_{L,k}+
 (D_{L\cup B}-D_{L,k})+(\mathcal R-D_{L\cup B}),
\]
measuring respectively the $k$-cell oracle floor, excess loss from the
specific extra labels $B$ deployed, and selection error within $L\cup B$
(valid under ties and baseline optima).

\subsection{Partial orthogonal cycles at an exact physical budget}
Subsampled Hadamard rows with randomized column signs are classical
sketching primitives \cite{tropp2011}, and full-cycle Hadamard designs
under bandit sums appear in \cite{rejwan2020}; here exact physical budgets
fix how many code rows each group receives in a single predetermined batch.
A group $G$ of $w$ labels in distinct blocks uses $w$ columns of the
Sylvester Hadamard matrix of order $H_G=2^{\lceil\log_2 w\rceil}$. Measuring
positive and negative parts of one code row costs two readings and $w$
settings, cancels $\mu_0$, and has conditional mean-zero noise of variance
at most $2\sigma^2$. Group $G$ receives $t_G$ differences consisting of
$\lfloor t_G/H_G\rfloor$ complete cycles plus $r_G=t_G\bmod H_G$ uniformly
sampled distinct rows, with independent Rademacher column signs across
columns and groups.

\begin{proposition}\label{prop:partial-cycle}
Condition on the groups and their counts before drawing code
randomness. For a cell $c$ let $q_c=\sum_{G\ni c}t_G>0$.
Multiply each incident difference by its sign at $c$ and average the
$q_c$ products to obtain $\widetilde\theta_c$. This estimate is
unbiased and satisfies
\begin{align}
 \mathbb E(\widetilde\theta_c-\theta_c)^2
 &\le {2\sigma^2\over q_c}
   +{1\over q_c^2}\sum_{G\ni c} B_G
                 \sum_{j\in G\setminus\{c\}}\theta_j^2,
 \label{eq:partial-risk}\\[-1mm]
 B_G&={r_G(H_G-r_G)\over H_G-1}\quad(H_G>1),
 \nonumber
\end{align}
with $B_G=0$ for singletons and equality under independent Gaussian reading
errors of variance $\sigma^2$.
\end{proposition}
\begin{proof}
Two distinct Hadamard columns have rowwise product summing to zero over a
complete cycle, so two distinct positions in a uniform row ordering have
product expectation $-1/(H_G-1)$, giving sum-of-products second moment
$r_G-r_G(r_G-1)/(H_G-1)=B_G$ over $r_G$ rows. Independent column and group
signs cancel cross terms, and conditional martingale-difference reading
errors contribute at most $2\sigma^2/q_c$ orthogonally to the signal term.
\end{proof}

When $w\ge2$ divides $K$, $L$ independent passes each cover all cells with
$t\le H_G$ distinct code rows per group ($Q=Lt$), and group partners are
drawn uniformly across other blocks and labels, the interference term in
\eqref{eq:partial-risk} averages to
\[
 { (H_G-t)(w-1)\over Q(H_G-1)(K-1)m}
                 \sum_{j\ne i}\sum_a\theta_{ja}^2 ,
\]
which is minimized at fixed $Q$ by the largest divisor $t\le H_G$ of $Q$.
In our experiments we also fit penalized regressions directly to these
measurement matrices.

\subsection{A guaranteed minimum number of visits}
For integer $P=T/2$ and $1\le C/P\le b$, split the $P$ contrasts between
$w_0=\lfloor C/P\rfloor$ and $w_0+1$ via $P_{w_0+1}=C-w_0P$ and
$P_{w_0}=P-P_{w_0+1}$ (omitting zero counts). At each active width $w$,
independently permute each block's labels into $m$ layers, partition each
layer's $K$ blocks into $\lceil K/w\rceil$ groups (padding the last group
with distinct blocks from earlier groups in that layer). Distribute the
$P_w$ contrasts so that group counts differ by at most one, then run
complete cycles and any remaining partial cycle within each group.
\begin{corollary}\label{cor:minimum-visits}
This design uses exactly $T$ readings and $C$ settings, and visits every
cell at least
\[
 q_{\min}=\sum_{w:P_w>0}
       \left\lfloor {P_w\over m\lceil K/w\rceil}\right\rfloor
\]
times. A schedule that completes each width-$w$ group before advancing,
with at most one partial final group per width, visits at most
$\sum_w w\lceil P_w/H_w\rceil$ distinct cells, where
$H_w=2^{\lceil\log_2 w\rceil}$.
\end{corollary}
\begin{proof}
Because $w\le K$, the padded group uses distinct blocks. Every cell belongs
to at least one of the $m\lceil K/w\rceil$ groups at each active width $w$,
receiving at least $\lfloor P_w/(m\lceil K/w\rceil)\rfloor$ visits, while
$2\sum_wP_w=T$ and $\sum_wwP_w=C$. Under group-by-group full cycles, each
started group covers at most $w$ distinct cells and every completed group
consumes $H_w$ contrasts.
\end{proof}

\paragraph{A population loss floor for missed positive labels.}
Draw $s$ active blocks uniformly without replacement, each with an
independent uniform positive optimum of amplitude $a\ge a_{\min}>0$ with
mean $\bar a$, and nonpositive effects elsewhere. For any instance-independent
schedule visiting fraction $p_i$ of labels in block $i$, any decoder that
selects nonbaseline labels only among visited cells incurs
\[
 \mathbb E[\mathcal R\mid\text{schedule}]
 \ \ge\ s\bar a\left(1-K^{-1}\sum_{i=1}^K p_i\right),
\]
and the number $M$ of missed active optima has conditional probability
generating function
\[
 \mathbb E[z^M\mid\text{schedule}]=
 {\binom Ks}^{-1}[t^s]\prod_{i=1}^K
       \{1+t[p_i+(1-p_i)z]\},
\]
bounding conditional success probability by
$\Pr\{M\le\lfloor\varepsilon/a_{\min}\rfloor\mid\text{schedule}\}$. At the
lower medium budgets, complete cycles visit at most $1{,}283$ of $2{,}048$
labels ($s=16$) and $2{,}675$ of $6{,}912$ labels ($s=24$); at
$s\bar a=3.36\varepsilon$, unvisited optima alone force expected loss at
least $1.255\varepsilon$ at $s=16$ and $2.059\varepsilon$ at $s=24$ under
either common decoder.

\section{Earlier feedback experiment}\label{sec:full-experiments}
ChatGPT (OpenAI) assisted with mathematical development and experimental implementation. The replication archive contains the locked study protocol, executable
policies and oracle, calibration splits, and pre-holdout hashes.

\subsection{Instances, physical noise, and the common decoder}
Effects are stored in units of $\varepsilon/s$. Each instance draws $s$
active blocks uniformly without replacement among $s^2$ blocks and assigns a
uniform best label in each active block value $3$ (total optimal reward
$3\varepsilon$). In the positive family all other labels are $0$; in the
signed family one additional uniform label per active block has value $-3$;
in the dense family the first $\lceil s/4\rceil$ active blocks assign
$\pm1.5$ to remaining labels and the rest follow the signed rule.

For a row of normalized mean $\mu$, the sum of $q$ fresh query--baseline
differences is $q\mu+E_q$. Under Gaussian noise,
$E_q=\sqrt{2\chi q}\,Z$ with $Z$ standard normal. Under Rademacher noise,
$E_q=2\sqrt\chi(U-V)$ with independent $U,V\sim\operatorname{Bin}(q,1/2)$.
The asymmetric law uses $E_q=2\sqrt\chi(U-q/5)-\sqrt\chi(2V-q)$ with
independent $U\sim\operatorname{Bin}(q,1/5)$ and $V\sim\operatorname{Bin}(q,1/2)$.
Each physical noise has valid proxy $\chi$. The three single-pair variances
are $2\chi$, $2\chi$, and $1.64\chi$, respectively. Each pair costs two readings.

The common decoder computes each cell's median across mixed-row means,
combines any singleton readings using pair-count weights, retains the
$\min(s^2,c_Us)$ blocks with largest absolute cell scores (stable index
ties), and fits weighted ridge regression over all labels of those blocks
with row weights $q_r/(2\chi)$ and penalty $c_\lambda$ times the median
design diagonal (LSMR tolerance $10^{-6}$, maximum $200$ iterations). In
each retained block it selects the maximum positive fitted label or
baseline. Finite iteration-limit outputs are scored as returned. Timeouts or nonfinite iterates return baseline and count as
failures. Ground truth enters only oracle generation and post-run scoring;
auxiliary diagnostics (active-block recall, shortlist recall, negative
selections, and singleton-only omission/ranking loss) are logged alongside
the ridge output.

\subsection{Exact allocation and feedback boundaries}
At degree $D$, $n_D=2Ds^2$ mixed rows have width $g=s/2$. For reference
budget $r$, set
\[
 Q=16s^2r,\quad q_0=\max\{1,\lfloor Q/(2n_D)\rfloor\},
 \quad R=Q-n_Dq_0,
\]
with $R\ge0$ divisible by $4$. The pilot uses $n_Dq_0$ pairs, the first and
second retests allocate $R/4$ pairs each across $n_D/4$ and $n_D/8$ nested
rows, and singleton validation allocates $R/2$ pairs as evenly as possible.
Every ordinary schedule spends
\[
 T=2Q,\qquad C_T=g(Q-R/2)+R/2,
\]
whereas mixed-only spends $C_T=gQ$. At the main grid's $D=8$ and $r=1$,
$R=0$ and all schedules coincide in a single batch.

Adaptive schedules rank mixed rows by absolute paired mean and form the
singleton shortlist from the two largest positive-direction scores in each
of the top $c_Us$ blocks. On its calibration split, fixed tuned selects its
mixed-row rule (uniform, random quarter, random eighth, or nested random)
and singleton rule (uniform coverage or random block/two-label shortlist);
fixed random uses nested random rows and a random shortlist.

Each batch commits its component action arrays and pair counts before any
observation in that batch is read, releasing results only at batch
boundaries. Role-indexed random streams pair identical components across
methods: two and four batches share pilot and first-retest data, two batches
and three-batch list-update share all mixed data, and three-batch row-update
and four batches share all mixed data.

\subsection{Calibration, seeds, and locked suites}
\begin{figure}[tp]
\centering
\includegraphics[width=.93\textwidth]{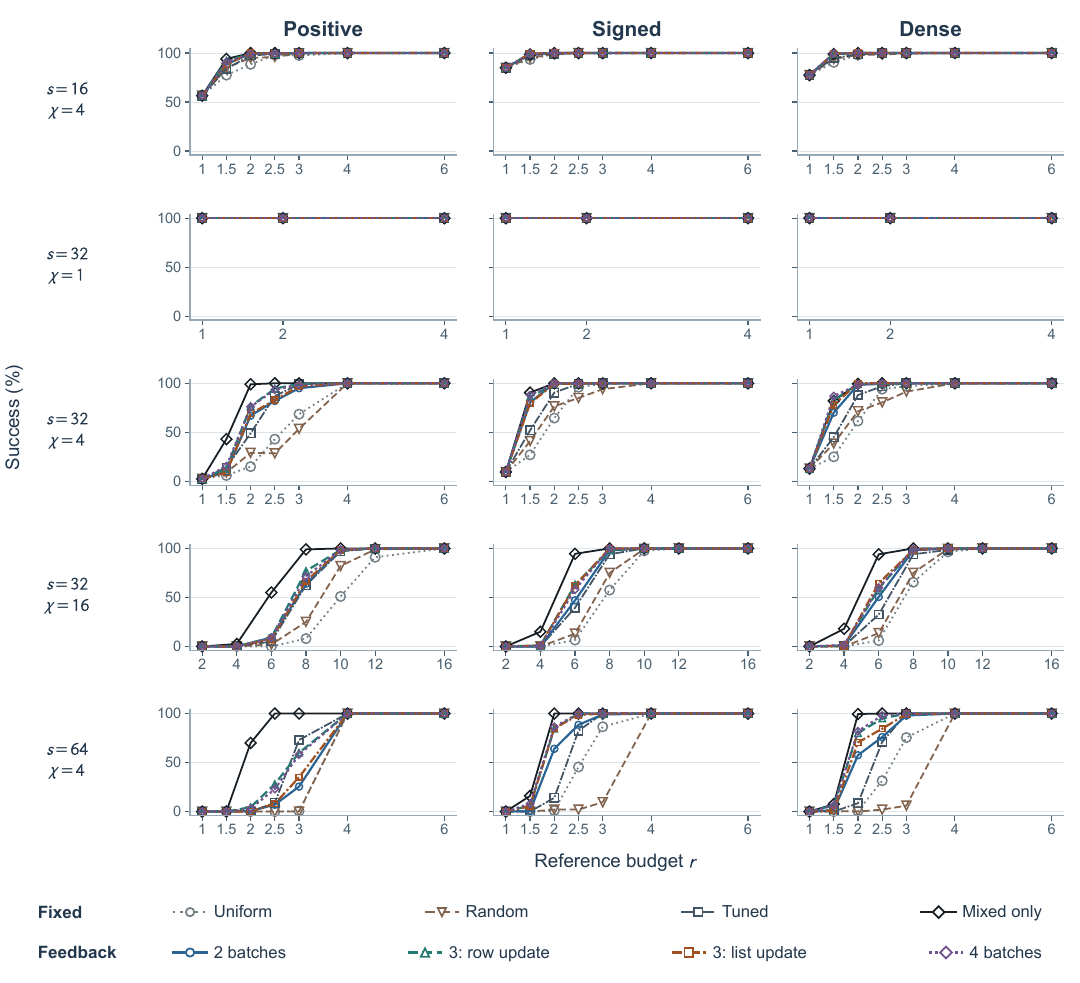}
\caption{Operational success across the $15$-panel budget grid in the
common-decoder study ($200$ paired instances per point, $148{,}800$ method
outcomes in total). Columns give the
positive, signed, and dense effect families; rows index $(s,\chi)$. The two
three-batch schedules update either the row ranking (3: row update) or the
singleton shortlist (3: list update).}
\label{fig:budget-grid}
\end{figure}

Figure~\ref{fig:budget-grid} reports success rates across the full $15$-panel
budget grid ($200$ paired instances per point, disjoint from the
$1{,}200$-instance primary holdout).
Root seed $4607302$ indexes independent streams by split, size, noise proxy,
family, replicate, degree, noise law, negative multiplier, and role; methods
are paired within each condition, while stress levels use distinct seeds.
Split~$21$ calibrates $(c_U,c_\lambda)\in\{2,4,8\}\times\{0.03,0.1,0.3\}$ by
mean regret across three conditions, three families, three budgets, $16$
instances, and four policies; split~$22$ calibrates the fixed allocation;
split~$23$ evaluates $80$ instances across $r\in\{1.5,1.75,\ldots,3.0\}$ to
pick the budget where two-batch success is closest to $0.65$ (Table~\ref{tab:suites}).

\begin{table}[!htbp]
\centering
\caption{Suites in the common-decoder study. Every method is evaluated on
every budget and instance within each split.}\label{tab:suites}
\begin{tabular}{lrr}
\toprule
Suite & Records & Split\\
\midrule
Decoder calibration & 15,552 & 21\\
Fixed-policy calibration & 3,456 & 22\\
Budget calibration & 560 & 23\\
Main grid & 148,800 & 30\\
Primary comparison & 9,600 & 31\\
Noise & 14,400 & 32\\
Layout degree & 4,800 & 33\\
Negative magnitude & 7,200 & 34\\
Larger scale & 2,400 & 35\\
Fresh timing & 240 & 36\\
\bottomrule
\end{tabular}
\end{table}

Main-grid budgets are $r\in\{1,2,4\}$ at $\chi=1$, $\{1,1.5,2,2.5,3,4,6\}$
at $\chi=4$, and $\{2,4,6,8,10,12,16\}$ at $\chi=16$. At $s=32,\chi=16$, the
noise ($200$ instances, $r\in\{4,8,12\}$, signed and dense), degree
($D\in\{4,8,16\}$, $100$ instances, $r\in\{4,8\}$), and negative-magnitude
($M\in\{1,10,100\}$, $200$ signed instances, $r\in\{4,8,12\}$) suites compare
fixed tuned, mixed only, two batches, and four batches; split~$35$ tests
$s=128,\chi=4$ on $100$ signed instances at $r\in\{2,3,4\}$.

\subsection{Inference and completeness}
For the primary four-minus-two success contrast at the preselected budget,
combining marginal $97.5\%$ Clopper--Pearson bounds $[l_{10},u_{10}]$ and
$[l_{01},u_{01}]$ on the two discordant binomial counts via
\[
 [l_{10}-u_{01},\ u_{10}-l_{01}]
\]
guarantees at least $95\%$ coverage by union bound (including zero counts),
alongside exact McNemar $p$-values and a $5{,}000$-resample paired bootstrap
interval for mean regret difference. Grid-panel $95\%$-success thresholds
report the smallest tested $r$ whose one-sided $1-0.05/M$ Clopper--Pearson
lower bound ($M$ method--budget cells in the panel) reaches $0.95$; scale,
mechanism, and stress suites are reported descriptively. All records,
component costs, trial keys, and hashes are verified on export and
stratified replay; single-worker timing measures computation and simulation
excluding physical oracle latency.

\subsection{Calibration, stress outcomes and runtime}
Decoder calibration minimized mean development regret ($0.6479\varepsilon$)
at $c_U=8$ and $c_\lambda=0.3$; fixed-policy calibration selected uniform
mixed retests with a random two-label shortlist; budget calibration selected
$r=2$ ($57/80$ two-batch successes). All $15$ budget panels, $93$ paired
grid contrasts, eight fixed-policy calibrations, and nine decoder
calibrations are included in the archive.

Table~\ref{tab:stress-compact} summarizes noise and pass-degree stress tests
at $s=32,\chi=16,r=8$. At $s=128$ and $\chi=4$ on signed instances, four
batches succeeded on $6/100$, $85/100$, and $100/100$ instances at
$r=2,3,4$; at $r=2$, mixed-only succeeded on $78/100$ while two batches
succeeded on $0/100$.

\begin{table}[!htbp]
\centering
\caption{Four-batch success rates (\%) at $s=32$, $\chi=16$, and $r=8$
($200$ instances per noise cell; $100$ per pass-degree cell).}\label{tab:stress-compact}
\begin{tabular}{lrr}
\toprule Condition & Signed & Dense\\\midrule
Gaussian noise &100.0&99.5\\
Rademacher noise &98.5&99.0\\
Asymmetric noise &100.0&100.0\\
Four passes &100.0&100.0\\
Eight passes &100.0&99.0\\
Sixteen passes &93.0&98.0\\
\bottomrule
\end{tabular}
\end{table}

All $187{,}200$ evaluation records, $19{,}568$ calibration records, and
$240$ timing records completed normally (LSMR stop code $2$), passing cost
audits across $665{,}456$ components and $52$ stratified hash-verified
replays. At the primary $s=64$ holdout, four batches reduced mean loss
relative to two batches by $0.1282\varepsilon$ (paired bootstrap 95\%
interval $[-0.1379,-0.1186]\varepsilon$). Across $30$ fresh single-thread
timing runs per policy (24-core host, 15.4~GiB RAM, Python 3.12.10, NumPy
2.3.5, SciPy 1.16.3, pandas 2.3.3), median (95th-percentile) compute-and-simulation
times were $0.153\,\text{s}$ ($0.162\,\text{s}$) for mixed only,
$0.195\,\text{s}$ ($0.202\,\text{s}$) for two batches, and
$0.196\,\text{s}$ ($0.209\,\text{s}$) for four batches.

\subsection{A noiseless obstruction for fixed ridge shrinkage}\label{sec:ridge-bias}
The practical decoder's positive ridge penalty is absent from the certified
theoretical decoder and introduces shrinkage bias when negative effects are
large.

\begin{proposition}[Ridge bias under legal aggregate rows]
For every fixed $\lambda>0$, ridge estimation followed by coordinatewise
choice against baseline can incur regret greater than $\varepsilon$
on a noiseless signed instance, even when its two nonzero cells are
known and its legal design has full column rank.
\end{proposition}
\begin{proof}
Embed two cells in distinct blocks ($s\ge2,b\ge2$), with all other effects
zero, row incidence matrix and effect vector
\[
 A=\begin{pmatrix}1&0\\0&1\\1&1\end{pmatrix},\qquad
 \theta=\varepsilon\begin{pmatrix}2\\-M\end{pmatrix},
\]
and noiseless baseline-differenced observations. With
$G=A^\top A=\left(\begin{smallmatrix}2&1\\1&2\end{smallmatrix}\right)$, the
ridge estimate $(G+\lambda I)^{-1}G\theta$ has first coordinate
\[
 \widehat\theta_1
 =\varepsilon\frac{6+4\lambda-\lambda M}{(2+\lambda)^2-1},
\]
and negative second coordinate for large $M$. Whenever $M>4+6/\lambda$,
coordinatewise maximization against baseline selects neither cell and incurs
regret $2\varepsilon$, whereas unpenalized least squares recovers $\theta$
exactly. Repeating $A$ $r$ times while scaling the penalty to $r\lambda$
(proportional to the design diagonal) leaves $\widehat\theta$ unchanged.
\end{proof}

Fixed positive shrinkage therefore cannot guarantee uniform PAC accuracy
across unbounded negative magnitudes.

\subsection{Independent penalty comparison}
In a noiseless, correctly specified fit with Gram matrix $G$, ridge bias
equals $-\lambda(G+\lambda I)^{-1}\theta$. After $720$ development outcomes
on split~$40$ ($10$ instances per magnitude), split~$41$ evaluated both
ridge and unpenalized regression after identical block selection on $200$
new signed instances across three negative multipliers, three budgets, and
four acquisition policies ($14{,}400$ outcomes; all observation hashes
matched and all solves completed). At the primary cell $(M,r)=(100,8)$ under
four batches, both fits achieved active-block recall $1$, while removing the
penalty lowered mean loss from $2.718\varepsilon$ to $0.356\varepsilon$ and
raised successes from $0/200$ to $200/200$ (simultaneous 95\% interval
$[95.67,100]$ points; exact two-sided McNemar $p=1.24\times10^{-60}$). Conversely,
at unit negative multiplier ($M=1$) on the same split, zero penalty reduced
four-batch success from $100.0\%$ to $94.5\%$.

\section{Pilot selection and packed-measurement studies}
\label{sec:new-experiments}
\begin{figure}[htbp]
\centering
\includegraphics[width=.94\textwidth]{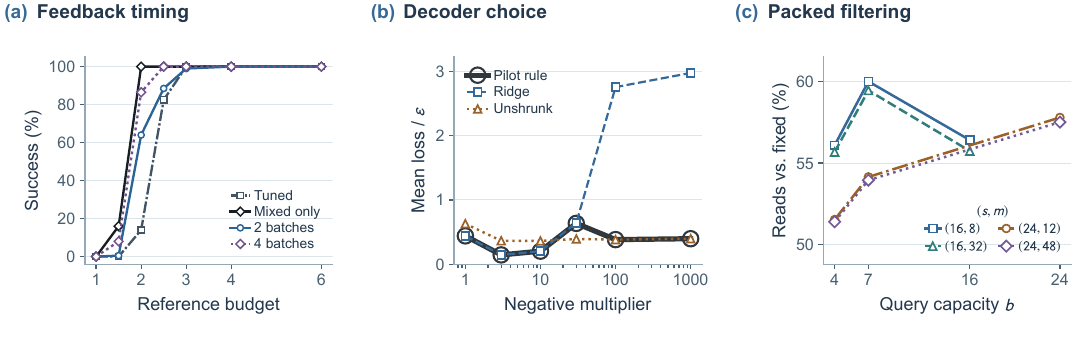}
\caption{Feedback timing, adaptive penalty selection, and packed two-stage
filtering. (a) Success rate across reference budgets $r$ ($s=64,\chi=4$,
signed instances, $200$ per budget). (b) Four-batch mean normalized loss at $s=32,\chi=16,r=8$
across negative multipliers $M\in[1,10^3]$ under fixed ridge, unpenalized
regression (Unshrunk), and the pilot rule ($160$ instances per multiplier).
(c) Mean within-pair ratio of two-stage to fixed-schedule readings for
each $(s,m)$ at $b=4,7,s$. Each point averages $384$ pairs across three
effect profiles.}
\label{fig:historical-panels}
\end{figure}

Figure~\ref{fig:historical-panels} summarizes feedback timing, decoder
shrinkage across negative multipliers, and two-stage orthogonal packing on
holdouts disjoint from Sections~\ref{sec:local-diagnostic}--\ref{sec:formal-studies}.

\subsection{A pilot rule with unchanged physical measurements}
From pilot means $y_i$ over $n$ mixed rows at $q_0$ pairs per row (in units
of $\varepsilon/s$), define the Gaussian-tail pilot statistic
\[
 S_{\rm pilot}
 =\frac{\max_i|y_i|}{\sqrt{4\chi\log(2n)/q_0}} ,
\]
evaluated at the regular pilot release boundary without altering any
subsequent measurement plan. On the $360$ paired development fits of
split~$40$ (three multipliers, three budgets, four policies, $10$
replicates), cutoffs $\{1,2,4,8,16\}$ and both constant rules were ranked by
mean normalized loss; cutoffs $8$ and $16$ tied at $0.6890625$, and the
prespecified larger-cutoff tie-break locked threshold $16$ (selecting zero
penalty when $S_{\rm pilot}>16$) before holdout generation.

Split~$42$ generated $160$ new signed instances at each
$M\in\{1,3,10,30,100,1000\}$ for $s=32,\chi=16$ and $r\in\{4,8,12\}$,
evaluating ridge, zero penalty, and the pilot rule on every shared plan
($34{,}560$ outcomes across $11{,}520$ plans on $960$ instances; all hashes
verified and all fits completed). Table~\ref{tab:gate} reports four-batch
successes at $r=8$. In the two prespecified primary contrasts (union-bounded
marginal $0.05/4$ Clopper--Pearson discordance intervals), the pilot rule
beat ridge at $M=100$ by $159$ wins to $0$ losses (simultaneous 95\%
interval $[0.924781,0.999961]$ as a fraction) and beat zero penalty at
$M=1$ by $13$ wins to $0$ losses ($[0.005374,0.150791]$).

\begin{table}[htbp]
\centering
\caption{Four-batch decoder successes out of $160$ new signed instances at
$s=32,\chi=16,r=8$ across negative multipliers $M$ (pilot cutoff calibrated
at $M\in\{1,10,100\}$ on split~$40$).}\label{tab:gate}
\begin{tabular}{rrrr}
\toprule $M$ & Ridge & Zero penalty & Pilot rule\\ \midrule
1 &160&147&160\\
3 &160&160&160\\
10&160&160&160\\
30&152&160&153\\
100&0&159&159\\
1000&0&160&160\\
\bottomrule\end{tabular}
\end{table}

\subsection{Orthogonal group experiment}
Using Lemma~\ref{lem:packing} at order $P=2^{\lceil\log_2 b\rceil}$ with
$x_1=\log H$, $x_2=H$, and $q_j=\lceil2\chi x_j/P\rceil$, two-stage
filtering retains cells whose normalized pilot estimate exceeds $0.75$
(returning baseline if retained cells exceed
$\min\{N,sm+s+\lceil N\exp(-0.1\log H)\rceil\}$), repacks survivors, and
maximizes stage-two estimates against baseline; the fixed comparator
measures all $N$ cells at $q_2$ repetitions, charging $2Pq_jG$ readings and
$Pq_jl$ settings for $l$ cells in $G$ groups. Root seed $4607318$ pairs
final Gaussian errors ($2\chi I/(Pq_j)$ by orthogonality, verified against
explicit physical row transforms) across common cells.

The $36$ panels cross $s\in\{16,24\}$, two label counts per $s$
($m\in\{8,32\}$ at $s=16$ and $m\in\{12,48\}$ at $s=24$), three capacities
$b\in\{4,7,s\}$ with $\chi=b$, and three effect profiles. The profiles
were signed, dense, and signed with the negative effect multiplied by $100$.
There were $128$ paired instances per panel, $4{,}608$ in total. Zero cap aborts
occurred, both arms succeeded on all $4{,}608$ instances, and mean
normalized loss was $0.01139$ for fixed measurement versus $0.01603$ for
filtering (within-pair cost ratios are averaged directly).

\paragraph{Checks of the geometric and numerical steps.}
Finite verification fixtures in the archive check the forest edge bound
across $127{,}202$ row-partition and source-set pairs, conditional occupancy
across $1{,}728$ labeled completions and $98$ joint-hit tests (confirming
why blocks exceeding $p/2$ exposed slots are counted deterministically),
and LP equivalence against row-submatrix enumeration on $30$ residual
systems. In an exact $K=4,m=2,b=2$ noiseless opposite-sign fixture (single-pass
pair collision probability $1/6$) with two- and four-pass filtering tables,
the half-mark rule misses the opposite pair on $21$ of $46{,}656$
equiprobable paths conditional on both cells reaching those tables.

\section{A controlled diagnostic for the residual measurement}
\label{sec:local-diagnostic}
Across eight independent seeds at $\varepsilon=1$, $K=64$, $m=4$, $s=b=8$,
$\sigma=1/8$, intercept $\mu_0=1.3$, and residual cap $k=2$, each seed placed
two strong cells ($+2\varepsilon,-2\varepsilon$) in distinct blocks alongside
four weak cells of magnitude $0.005\varepsilon$, evaluating both sign
orientations of the strong pair under shared noise and weak effects ($16$
coupled instances per layout).

Each layout used four balanced passes, eight groups per layer, and $16$
fresh Gaussian row--baseline pairs across $128$ rows ($256$ candidate cells,
$T=4{,}096$, $C_T=16{,}384$). Conditioning the repeated-layout control on
repeating a pass in which the strong pair collides isolates structural
nonidentifiability, whereas the fresh control draws four independent
unconditioned passes. On each dataset we compared the incident-row median
with the convex residual LP of Section~\ref{sec:res-cert} (retaining the two
largest absolute entries before maximizing against baseline) using top-$8$
residual radius
\[
 8\left(0.02+\sqrt{4\sigma^2\log(2\cdot128/0.05)/16}\right).
\]

\begin{table}[!htbp]
\centering
\caption{Coupled residual diagnostic ($8$ seeds $\times$ $2$ sign
orientations per row; $T=4{,}096$, $C_T=16{,}384$).}\label{tab:local}
\begin{tabular}{llrr}
\toprule Layout & Fit & Successes & Mean loss / $\varepsilon$\\\midrule
Repeated & Median & 0/16 & 2.00875\\
Repeated & LP & 0/16 & 2.00875\\
Fresh & Median & 16/16 & 0.00875\\
Fresh & LP & 16/16 & 0.00875\\
\bottomrule\end{tabular}
\end{table}

On the repeated layout, both decoders returned identical assignments across
the two sign orientations ($0/16$ successes); on fresh passes, both
separated the canceling pair on all $16/16$ instances (Table~\ref{tab:local}).
Additional archive fixtures verify the length-two path bound on $8{,}445$
subsets, largest-row factorial moments on enumerated passes, and $\ell_1$
error bounds on $24$ small certified systems.

\section{Additional paired studies}\label{sec:formal-studies}
Throughout this section $\varepsilon=1$; operational success requires normal
completion with $\mathcal R_\theta(\widehat x)\le1$ (the archive also logs
realized baseline loss after failed checks or solves). Each paired unit
evaluates all compared methods on a shared instance, coupling identical
plans to identical noise streams while drawing fresh noise for distinct
pairs within a method.

\subsection{Prespecified finite-size comparisons}
Four primary comparisons were locked prior to evaluation ($2{,}048$ paired
instances each, $\chi=4$, $h=\log20$). Three residual comparisons use $s=6$,
$K=36$, $m=3$, $b=6$, $d=3$, public cap $k=6$, six alternating-sign strong
cells in distinct blocks with magnitudes uniform on $[0.704,0.896]$, and
total weak magnitude at most $0.02$: (1) top-$dk$ convex LP versus incident
median on identical fresh matrices and responses; (2) holding the convex LP
fixed, best of four layouts ranked lexicographically by
$(u(1+\sqrt{2kv}),u,\text{draw index})$ versus the first draw on fresh
layouts; and (3) the same four-draw ranking when each draw independently uses
fresh, partially repeated, or fully repeated passes with probabilities
$0.50$, $0.35$, and $0.15$ (strict sufficient-certificate outcomes are
logged separately).

The fourth comparison evaluates complete policies from raw observations at
$s=8$, $K=64$, $m=4$, $b=6$, $\chi=4$, where each of the eight active blocks
contains $+a$ and $-a$ with $a$ uniform on $[0.88,1.12]\times0.42$. The
adaptive policy runs two pilot passes ($2$ pairs/row), selects $12$
shortlist cells by absolute incident median, and in batch two jointly runs
$8$ singleton pairs per shortlisted cell plus three masked residual passes
($2$ pairs/row), decoding with residual cap $k=4$ ($T=1{,}072$,
$C_T=2{,}584$). The fixed comparator spends the exact same $(T,C_T)$ on a
predetermined signed Hadamard design decoded by ridge regression.

For $n$ residual rows, $r=\min(n,dk)$, $q$ pairs/row, and public weak bound
$W$, the primary LP radius equals $rW$ plus the smaller of
\begin{align*}
 &r\sigma\sqrt{4(\log(2n)+h)/q},\\
 &\sigma\sqrt{4r\{\log\tbinom nr+r\log2+h\}/q},
\end{align*}
with $W=0$ in the $s=8$ full-policy comparison (actual signal tails are
logged). Repetition counts were fixed on two disjoint development splits
($32$ median-only and $48$ single-layout zero-feasibility instances) without
between-policy tuning. Sensitivity panels vary dimensions, pass counts,
budgets, signs, dynamic range, noise law, residual caps, fixed-truth
replicates, high confidence ($h=2s$), multiscale cap exits ($d_0=8$,
precision factor $8$, radius $d/128$, $\lceil(h+\log64)/\log2\rceil$ strict
tries), and GPU/CPU compute profiles.

\subsection{Matched medium-scale designs}
Four strata cross $(s,m,b)\in\{(16,8,12),(24,12,18)\}$ with low and high
budgets ($\chi=4$, Gaussian noise, one $+a$ and one $-a$ per active block
with $a$ uniform on $[0.88,1.12]\times3.36/s$). The two-batch adaptive plan
takes two pilot passes, shortlists $3s/2$ cells, and runs singleton
validation plus three masked residual passes at $(q_0,q_V,q)=(2,8,2)$ (low)
or $(4,16,4)$ (high) with public cap $s/2$; a signed-residual variant
randomizes fresh residual signs while sharing pilot, shortlist, geometry,
and noise streams.

Fixed plans allocate $P=T/2$ signed contrasts across the two neighboring
widths of $C_T/P$, comparing complete groupwise Hadamard cycles against
four-row shortened cycles. Under the global ridge decoder, all released
pair-weighted differences solve $\min_z\|Dz-y\|_2^2+\lambda\|z\|_2^2$ at a
shared absolute penalty $\lambda=0.1C_T/(Km)$ (plus a final-readings-only
ablation). Under the block-refit decoder, column-normalized scores
$|(D^\top y)_j|/\|D_{\cdot j}\|_2$ ($0$ if unvisited) rank blocks by the sum
of their top two squared cell scores, retain the top $cs$ blocks, and refit
ridge there with penalty $\gamma C_T/(Km)$; both decoders output positive
maxima in at most the top $s$ positive blocks.

To test the effect of the residual radius, we drew $8{,}192$ standardized
Gaussian vectors for each of the two row shapes. For each shape we used
the $7{,}833$rd ordered value of the largest-$r$ absolute sum $J_r$.
The probability that this value falls below the population 95th percentile
is $\Pr\{\operatorname{Bin}(8192,0.95)\ge7833\}\le0.005$. Multiplying
by $\sigma\sqrt{2/q}$ gives simultaneous upper bounds for both noise-only
percentiles with calibration confidence of at least $0.99$.
Nonzero residual LPs were solved in HiGHS
($30\,\text{s}$ limit; $8{,}192\times2{,}400$ cap in medium strata and
$1{,}800\times1{,}800$ in the primary suite) with independent primal--dual
verification.

Sixteen development instances per stratum fixed the short-cycle row cap at
$4$ (from $\{1,2,4,8\}$) and $(c,\gamma)=(1,0.1)$ (from
$\{1,2,4\}\times\{0.03,0.1\}$) before generating $256$ evaluation pairs per
stratum, followed by $192$ independent pairs per stratum comparing complete
cycles, four-row cycles, and the untuned minimum-visit schedule of
Corollary~\ref{cor:minimum-visits} under both common decoders.

\paragraph{Reconstructing the article figures.}
Article Fig.~1c pairs two- and four-batch outcomes across the $1{,}200$
primary $s=64$ instances ($750$ joint successes, $320$ four-batch-only, $29$
two-batch-only, $101$ joint failures). Article Fig.~2 plots $U_B$ at
$s=256$. Article Fig.~3b,c plots all $2{,}048$ full-policy $s=8$ pairs at
$\varepsilon=1$, using $D_L(\theta)$ on the horizontal axis of Fig.~3c while
$D_{L,4}$ enters the three-term loss decomposition. Article Fig.~4a,b
averages cumulative distinct-label coverage and terminal visit-tail counts
across the $192$ hash-verified layouts per schedule in the low-budget $s=24$
stratum; Fig.~4c plots the $256$-pair medium-scale adaptive-data decoder
comparison; Fig.~4d plots paired block-refit loss differences relative to
minimum visits on the $192$ low-budget $s=24$ instances.

\subsection{Inference and reporting}
Within a family of $M$ prespecified paired comparisons ($M=4$ primary,
$M=28$ medium-scale, and $M=16$ minimum-visit comparisons), subtracting
opposite endpoints of marginal Clopper--Pearson intervals at tail level
$0.05/(4M)$ yields simultaneous $95\%$ intervals for discordant risk
differences $p_{10}-p_{01}$, paired with Holm-adjusted exact two-sided
McNemar tests. Secondary means and failure-mode breakdowns are descriptive.

\subsection{Complete accounting and primary outcomes}
The $342$-condition checkpoint comprises $34{,}064$ paired units and
$188{,}528$ method or diagnostic records (Table~\ref{tab:status-5080});
Table~\ref{tab:main-5080} reports success counts and discordances for the
four primary comparisons.

\begin{table}[htbp]\centering\caption{Accounting across the $342$-condition suite. Exits count declared dimension, retry, or cap terminations.}\label{tab:status-5080}
\begin{tabular}{lrrr}\toprule Panel & Units & Records & Exits\\\midrule
Primary residual & 6,144 & 61,440 & 12,288\\
Primary full policy & 2,048 & 14,336 & 0\\
Residual sensitivity & 5,280 & 52,800 & 10,368\\
Full-policy sensitivity & 2,208 & 15,456 & 0\\
Fixed instance & 6,144 & 12,288 & 0\\
Confidence & 4,608 & 4,608 & 0\\
Decoder frontier & 3,072 & 12,288 & 0\\
Geometry & 1,920 & 1,920 & 240\\
Multiscale & 1,024 & 4,096 & 1,030\\
Scale profile & 80 & 80 & 0\\
Earlier policies & 1,536 & 9,216 & 0\\
\bottomrule\end{tabular}\end{table}
\begin{table}[htbp]\centering\caption{Four prespecified primary comparisons ($2{,}048$ paired instances per row; percentage-point differences and simultaneous 95\% intervals).}\label{tab:main-5080}
\small\setlength{\tabcolsep}{4pt}
\begin{tabular}{@{}p{.31\linewidth}rrrrr@{}}\toprule Comparison & \shortstack[r]{A\\successes} & \shortstack[r]{B\\successes} & A only & B only & \shortstack[r]{Difference\\(95\% interval)}\\\midrule
Convex versus median, fresh & 526 & 1,029 & 188 & 691 & $-24.56\ [-29.14,-19.85]$\\
Ranked versus first, fresh & 557 & 530 & 231 & 204 & $+1.32\ [-2.45,5.08]$\\
Ranked versus first, mixture & 445 & 435 & 220 & 210 & $+0.49\ [-3.26,4.24]$\\
Adaptive convex versus fixed orthogonal & 26 & 1,861 & 5 & 1,840 & $-89.60\ [-91.54,-87.17]$\\
\bottomrule\end{tabular}\end{table}
In the geometry panel, exact $(u,v)$ statistics were computed for all four
draws in $1{,}680$ of $1{,}920$ units. Under the strict pair and
third-column tests, the first-draw and four-draw rules each accepted $240$
units, including singleton-width controls. Across the $2{,}048$
hash-verified $s=8$ full-policy instances, the true residual contained $7$
to $16$ nonzero cells against public cap $k=4$ (mean tail mass $2.86$
outside the top four residual cells); $1{,}114$ of the $1{,}120$
zero-feasible fits had $D_L(\theta)>1$, and $467$ of $2{,}048$ instances
had four-cell oracle floor $D_{L,4}>1$, decomposing mean loss $2.169$ into
$0.636$ (cap $D_{L,4}$), $1.532$ (deployed extra cells), and $0.001$
(within-set selection).

\subsection{Medium holdouts and the independent coverage study}
Lower $(T,C_T)$ budgets were $(3{,}904,20{,}528)$ at $s=16$ and
$(8{,}256,69{,}192)$ at $s=24$, and doubled at the higher budgets.
Tables~\ref{tab:medium-5080} and~\ref{tab:coverage-5080} report successes
and mean normalized losses across the $256$-pair medium-scale evaluation and
the $192$-pair coverage holdout; Table~\ref{tab:radius-5080} gives zero-fit
counts before and after Gaussian radius calibration.

\begin{table}[htbp]\centering
\caption{Matched medium-scale evaluation: successes out of $256$ paired instances and mean normalized loss in parentheses. Unsigned and signed adaptive plans share pilots and shortlists; each decoder uses a shared penalty across designs within a stratum.}\label{tab:medium-5080}
\small\setlength{\tabcolsep}{4pt}
\begin{tabular}{lrrrr}\toprule
Acquisition and decoder & $s=16$, low & $s=16$, high & $s=24$, low & $s=24$, high\\\midrule
Adaptive, original convex & 0 (2.64) & 0 (1.87) & 0 (2.78) & 0 (2.07)\\
Adaptive, Gaussian-calibrated convex & 0 (2.39) & 0 (2.01) & 0 (2.55) & 0 (2.22)\\
Adaptive, final readings, ridge & 3 (1.73) & 91 (1.08) & 0 (2.02) & 27 (1.34)\\
Adaptive, all readings, ridge & 25 (1.44) & 230 (0.64) & 6 (1.61) & 226 (0.75)\\
Adaptive, all readings, block refit & 221 (0.65) & 256 (0.18) & 214 (0.81) & 256 (0.21)\\
Signed residual, ridge & 15 (1.43) & 234 (0.64) & 5 (1.60) & 226 (0.74)\\
Signed residual, block refit & 221 (0.64) & 256 (0.19) & 209 (0.80) & 256 (0.19)\\
Fixed complete cycles, ridge & 18 (1.58) & 234 (0.54) & 0 (2.17) & 73 (1.20)\\
Fixed complete cycles, block refit & 17 (1.61) & 236 (0.52) & 0 (2.17) & 65 (1.22)\\
Fixed shortened cycles, ridge & 5 (1.65) & 254 (0.38) & 0 (1.91) & 200 (0.83)\\
Fixed shortened cycles, block refit & 243 (0.54) & 256 (0.03) & 228 (0.72) & 256 (0.06)\\
\bottomrule\end{tabular}
\end{table}
\begin{table}[htbp]\centering
\caption{Independent coverage holdout: successes out of $192$ paired instances and mean normalized loss in parentheses under locked budgets and decoder parameters.}\label{tab:coverage-5080}
\small\setlength{\tabcolsep}{4pt}
\begin{tabular}{lrrrr}\toprule
Acquisition and decoder & $s=16$, low & $s=16$, high & $s=24$, low & $s=24$, high\\\midrule
Fixed complete cycles, ridge & 15 (1.60) & 179 (0.52) & 0 (2.21) & 58 (1.23)\\
Fixed complete cycles, block refit & 12 (1.62) & 182 (0.50) & 0 (2.22) & 54 (1.25)\\
Fixed shortened cycles, ridge & 6 (1.65) & 192 (0.39) & 0 (1.92) & 161 (0.79)\\
Fixed shortened cycles, block refit & 185 (0.54) & 192 (0.02) & 176 (0.71) & 192 (0.06)\\
Fixed minimum-visit design, ridge & 5 (1.60) & 192 (0.24) & 0 (1.97) & 186 (0.63)\\
Fixed minimum-visit design, block refit & 184 (0.49) & 192 (0.02) & 171 (0.71) & 192 (0.04)\\
\bottomrule\end{tabular}
\end{table}
\begin{table}[htbp]\centering\caption{Zero-fit counts out of $256$ instances for the original deterministic and Gaussian-calibrated residual LP radii.}\label{tab:radius-5080}
\begin{tabular}{lrr}\toprule Stratum & Original radius & Calibrated radius\\\midrule
16, low & 181 & 0\\
16, high & 3 & 0\\
24, low & 238 & 0\\
24, high & 0 & 0\\
\bottomrule\end{tabular}\end{table}

At the lower budgets, complete cycles visited $52.9\%$ of labels at $s=16$
and $35.2\%$ at $s=24$ on average. On identical adaptive readings, block
refitting exceeded global ridge by $+76.56$ points at $s=16$ low
(simultaneous 95\% interval $[63.88,84.65]$) and $+81.25$ points at $s=24$
low ($[69.10,88.52]$), whereas randomizing final-batch residual signs yielded
no resolved gain under block refitting. In the independent low-budget $s=24$
coverage holdout under block refitting, minimum visits beat complete cycles
by $+89.06$ points ($[76.59,94.95]$) and differed from four-row cycles by
$-2.60$ points ($[-15.42,10.42]$).

Among $1{,}024$ multiscale first-layout convex runs, $511$ returned
normally, $127$ were infeasible, $383$ hit the dimension cap, $1$ returned a
solver error, and $2$ exceeded the screen cap; under the strict certificate
policy, all $511$ positive-residual runs exhausted their retry allowance,
leaving only $k=0$ paths completing normally. On the $2{,}097{,}152$-cell
scale profile ($10$ runs), RTX~5080 Laptop GPU row-sum and incident-median
kernels took $32.2\,\text{ms}$ and $33.2\,\text{ms}$ median warm time, while
exact $(u,v)$ certification and CPU LP solves hit their declared size caps.

\appendix

\section{The unrestricted-product CHO-PE baseline}\label{sec:cho}
This appendix treats unrestricted capacity $b=K$ for general $K\ge s\ge1$
and $M=m+1\ge2$ labels per block (including baseline), giving a constructive
scalar-query upper bound for full-product actions.

\begin{proposition}[Legal Hadamard packing and parallel elimination]\label{prop:cho}
With a supplied valid aggregate proxy $\sigma^2$, CHO-PE is $(\varepsilon,\delta)$-PAC and has the pathwise scalar-query cap
\[
 N_H\le C M\left(s+\frac{\sigma^2s^2}{\varepsilon^2}\right)
 \log(es/\delta).
\]
This count applies on all hash and noise paths without support-recovery or minimum-effect assumptions.
\end{proposition}
\begin{proof}
Set $B_H=2^{\lceil\log_2(16s)\rceil}$ and let $R_H$ be the smallest odd
integer at least $(128/9)\log(2s/\delta)$. Draw $R_H$ independent
pairwise-independent hash functions $h_r:[K]\to[B_H]$ and sign maps
$g_r(i)\in\{-1,1\}$ once before observing data. Each active block collides
with another active block in table $r$ with probability at most $1/16$, so
Hoeffding's inequality and a union over at most $s$ active blocks yield an
initial seed event $\mathcal H$ of probability at least $1-\delta/2$ on
which every active block is isolated in at least $3R_H/4$ tables across all
subsequent rounds and candidate lists.

Given requested labels $a_i$ and Sylvester row $k$ of order $B_H$, place
$a_i$ in the positive query and $0$ in the negative query when
$g_r(i)H_{k,h_r(i)}=1$, and reverse them otherwise (baseline requests place
$0$ in both), which is legal at $b=K$. Their mean difference equals
$(H\beta)_k$ with $\beta_b=\sum_{i:h_r(i)=b}g_r(i)\theta_{i,a_i}$; averaging
$q$ fresh pairs and multiplying by $B_H^{-1}H^\top$ gives each bucket proxy
$2\sigma^2/(qB_H)$, so multiplying by $g_r(i)$ estimates any isolated block.

Starting from $M$ candidates per block, retain $\lceil m_\ell/2\rceil$ per
round under
\[
 \begin{aligned}
 m_\ell&=\left\lceil M/2^{\ell-1}\right\rceil,&
 L&=\lceil\log_2M\rceil,\\
 \gamma_\ell&=\frac{\varepsilon}{4s}(3/4)^{\ell-1},&
 \rho_\ell&=\frac{\delta}{6s2^\ell},
 \end{aligned}
\]
with baseline estimates fixed to $0$, deterministic tie-breaking, and
\[
 q_\ell=\max\left\{1,
 \left\lceil\frac{64\sigma^2}{B_H\gamma_\ell^2}
 \left(\log2+R_H^{-1}\log(2/\rho_\ell)\right)\right\rceil\right\}.
\]
Conditional on $\mathcal F_{\ell-1}$ and $\mathcal H$, an error exceeding
$\gamma_\ell/2$ in the table-median estimate requires at least $R_H/4$
isolated tables to err in the same direction, which has conditional
probability
\[
 \Prob\{|\widehat\theta_{i,a}-\theta_{i,a}|>\gamma_\ell/2
       \mid\mathcal F_{\ell-1}\}
 \le2\exp\!\left(R_H\log2-
 \frac{q_\ell B_H\gamma_\ell^2R_H}{64\sigma^2}\right)
 \le\rho_\ell
\]
(and zero for $\sigma=0$). In any active block, its current true maximizer
is underestimated by more than $\gamma_\ell/2$ with probability at most
$\rho_\ell$; otherwise, a drop exceeding $\gamma_\ell$ in the best retained
true value requires at least $m_\ell/2$ retained candidates to overestimate
by more than $\gamma_\ell/2$, which Markov's inequality bounds by
$2\rho_\ell$ without assuming independence across candidates. Unioning over
rounds and active blocks adds at most $3s\sum_\ell\rho_\ell\le\delta/2$ to
$\Prob(\mathcal H^c)\le\delta/2$, while $\sum_\ell\gamma_\ell\le\varepsilon/s$
keeps total regret at most $\varepsilon$.

Executing all slots deterministically uses
$2R_HB_H\sum_{\ell=1}^L m_\ell q_\ell$ readings; since
$m_\ell\le2M2^{-(\ell-1)}$, $(1/2)(4/3)^2=8/9<1$, $B_H=\Theta(s)$, and
$R_H=\Theta(\log(es/\delta))$, the stated cap holds pathwise (using
$O(KR_H)$ hash storage and $O(KN_H)$ coordinate submissions, with $\sigma_K^2$
substituted when variance scales with $K$).
\end{proof}

\subsection{An equal-task comparison}
In the Gaussian subclass with $K=s\ge2$, $b=K$, $0<\delta<1/16$, and fixed
$\sigma/\varepsilon\ge1$, baseline normalization gives ambient linear
dimension $D_s=1+s(M-1)$ and $|\mathcal X|=M^s$. For fixed
$s,\delta,\sigma/\varepsilon$, Proposition~\ref{prop:cho} gives upper bound
$O(M)$, matching the dependence on $M$ in the multitask lower bound of
Maiti et al.~\cite{maiti2026}. Substituting the same action set into their
general finite-action upper bound (Theorem~6) gives
\[
 O\left(\frac{\sigma^2D_s}{\varepsilon^2}
 \left[\log(M^s/D_s)+\log(1/\delta)\right]\right)=O(M\log M).
\]
The capacity-constrained feedback separation follows from
Theorem~\ref{thm:uniform} under \eqref{eq:benchmark}. Separately, Malek et
al.~\cite{malek2023} establish instance-dependent elimination bounds for
additive causal bandits under causal-graph intervention assumptions distinct
from the worst-case capacity-constrained formulation in \eqref{eq:benchmark}.

\renewcommand{\bibfont}{\small}
\setlength{\bibsep}{2pt plus 1pt minus 1pt}
\bibliographystyle{elsarticle-num}
\bibliography{references}